\documentclass[]{tJOT2e}

\usepackage{graphicx}
\usepackage{amsmath}
\usepackage{hyperref}
\usepackage{color}
\usepackage{comment}

\newcommand{\bs}{\boldsymbol}

\begin{document}
\doi{10.1080/14685248.2015.1047497}
\issn{1468-5248}
 \jvol{16} \jnum{10} \jyear{2015}

\markboth{Taylor \& Francis and I.T. Consultant}{Journal of Turbulence}

\title{Height-dependence of spatio-temporal spectra of wall-bounded turbulence -- LES results and model predictions}

\author{Michael Wilczek$^{\rm a,b}$$^{\ast}$\thanks{$^\ast$Corresponding author. Email: michael.wilczek@ds.mpg.de}, Richard J.A.M. Stevens$^{a,c}$ and Charles Meneveau$^a$\\
$^a$Department of Mechanical Engineering, Johns Hopkins University, Baltimore, Maryland 21218, USA.\\
$^b$Max Planck Institute for Dynamics and Self-Organization, D-37077 G\"ottingen, Germany.\\
$^c$Department of Science and Technology and J.M. Burgers Center for Fluid Dynamics, University of Twente, P.O. Box 217, 7500 AE Enschede, The Netherlands.
}

\maketitle

\begin{abstract}

Wavenumber-frequency spectra of the streamwise velocity component obtained from large-eddy simulations are presented. Following a recent paper  [Wilczek et al., J. Fluid. Mech., 769:R1, 2015] we show that the main features, a Doppler shift and a Doppler broadening of frequencies, are captured by an advection model based on the Tennekes-Kraichnan random-sweeping hypothesis with additional mean flow. In this paper, we focus on the height-dependence of the spectra within the logarithmic layer of the flow. We furthermore benchmark an analytical model spectrum that takes the predictions of the random-sweeping model as a starting point and find good agreement with the LES data. We also quantify the influence of LES grid resolution on the wavenumber-frequency spectra.

\begin{keywords}
wall-bounded turbulence, space-time correlations, turbulence modeling, large-eddy simulations
\end{keywords}

\end{abstract}

\section{Introduction}

Wall-bounded flows exhibit a rich spatio-temporal structure with high- and low-speed streaks superposed with turbulent fluctuations (see, e.g. \cite{smits11arf,jimenez13pof} and references therein). Among the most fundamental statistical objects to characterize such flows is the space-time correlation, or equivalently, the wavenumber-frequency spectrum of the streamwise velocity. Understanding spatio-temporal correlations is not only interesting from a fundamental fluid dynamics research point of view, but has also implications for the study of atmospheric boundary layers with potential applications in the field of wind energy research.

Wavenumber spectra for wall-bounded flows characterizing the spatial correlations have been subject to intense theoretical, experimental and numerical research for a long time (see, e.g., \cite{perry82jfm,perry86jfm,kim87jfm,smits11arf,jimenez13pof}), but only recently measurements of the wavenumber-frequency spectra from a wall-bounded flow have been obtained \cite{lehew11exf}, giving insights into spatio-temporal correlations.

Following up on a recent work \cite{wilczek15jfm}, we present here further results on wavenumber-frequency spectra of high-Reynolds number data obtained from large-eddy simulation (LES) of wall-bounded turbulence. We are especially interested in characterizing the logarithmic layer of the flow and hence focus here on the discussion of the height-dependence of the spatio-temporal spectra in this layer. To this end, wavenumber-frequency spectra of the streamwise velocity fluctuation as function of the streamwise wavenumber and frequency as well as of the spanwise wavenumber and frequency are discussed. We furthermore evaluate the influence of LES grid resolution.

In characterizing the spatio-temporal correlations of turbulent fluctuations in wall-bounded flows, advection of the fluctuations by the mean flow as well as by large-scale eddies turn out to be the dominant effects. The latter effect gives rise to Eulerian time decorrelations and led Kraichnan and Tennekes to the introduction of the random-sweeping hypothesis \cite{kraichnan64pof,tennekes75jfm}. In the framework of this hypothesis smaller-scale velocity fluctuations are considered as passively advected by large-scale energy-containing eddies in a random fashion. The random-sweeping hypothesis plays an important role in various models for space-time correlations \cite{he06pre,zhao09pre,wilczek12pre}, but has also been used to quantify corrections to Taylor's frozen turbulence hypothesis, as discussed, e.g., in \cite{lumley65pof,wyngaard77jas,george89afm,alamo09jfm,wilczek14npg}.

In a recent paper, we explored a simple advection model based on the Kraichnan-Tennekes random-sweeping hypothesis with additional mean flow for the space-time correlations of wall-bounded flows \cite{wilczek15jfm}, which we will briefly summarize. In the present paper we further substantiate the validity of this random-sweeping advection model with mean flow in characterizing space-time correlations of wall-bounded turbulent flow and present tests of its predictions for various heights.

The random-sweeping model with mean flow predicts the wavenumber-frequency spectrum to be a product of the wavenumber spectrum and a Gaussian frequency distribution, which is parameterized by a Doppler shift related to the mean velocity and a Doppler broadening related to random-sweeping effects. The precise shape of the wavenumber spectrum as well as of the Doppler shift and broadening as a function of wall distance, however, remain yet to be determined. In \cite{wilczek15jfm} we obtained analytical model parameterizations for these terms. For the wavenumber spectrum we blend two phenomenological model spectra for the low and high wavenumbers, respectively. The height-dependence  of the Doppler shift and broadening terms are parameterized with the help of logarithmic laws for the mean and fluctuating velocity. Here, we further benchmark this model with particular focus on the height-dependence of the spectra as well as on the influence of grid resolution.

The article is structured as follows. We first present wavenumber-frequency spectra evaluated from LES for various heights in the logarithmic layer of the flow as function of the streamwise wavenumber and frequency as well as of the spanwise wavenumber and frequency. These results are then compared to the predictions of a simple advection model. Furthermore, comparisons to the analytical model are presented. Finally, we assess the influence of LES grid resolution on our results, before we conclude.

\section{Wavenumber-frequency spectra from LES at various heights}

The numerical data presented in this paper are obtained from LES of wall-modeled half-channel flow with rough walls ($z_0/H=10^{-4}$)\citep{albertson99wrr,porteagel00jfm,bou05,ste14d}. In horizontal directions a pseudo-spectral discretization with periodic boundary conditions is implemented. The vertical direction is discretized with second-order finite differences. To model the sub-grid scale stresses we employ the Lagrangian scale-dependent model \citep{bou05}. In absence of viscous scales we cannot associate a well-defined Reynolds number to our simulation but the LES is meant to represent high-Reynolds number flow.

The computational domain $L_x/H \times L_y/H \times L_z/H = 4\pi \times 2\pi \times 1$ is resolved with $1024 \times 512 \times 256$ grid points. The simulation is run for $8.2 \times 10^4$ time steps at a fixed time step of $2.5 \times10^{-5} H/u_*$ in the statistically stationary regime. We refer the reader to \cite{ste14d,wilczek15jfm} for further details; the data set used here corresponds to case D2 in \cite{ste14d}.

To characterize the spatio-temporal correlations of the streamwise velocity component $u$ in horizontal planes at various heights we study the wavenumber-frequency spectrum $E_{11}(\bs k,\omega;z)$, which is a function of the horizontal wavenumbers $k_1$ and $k_2$, the frequency $\omega$ as well as the distance from the wall, $z$. For presentation purposes, as well as to increase statistical convergence, we consider the projections
\begin{align}\label{eq:projection}
 E_{11}(k_1,\omega;z) &=  \int\!\mathrm{d}k_2 \, E_{11}(\bs k,\omega;z) \\
 E_{11}(k_2,\omega;z) &=  \int\!\mathrm{d}k_1 \, E_{11}(\bs k,\omega;z) \, .
\end{align}
Streamwise wavenumber-frequency spectra for various heights are shown in the left panels of figure \ref{fig:stream_wf_spec_vs_model}. The spectra exhibit a Doppler shift of frequencies towards higher wavenumbers. This spectral tilt is an immediate consequence of the mean flow advection. Smaller-scale velocity fluctuations are experiencing a more pronounced Doppler shift. As the mean velocity increases with increasing distance to the wall, the Doppler shift is less pronounced closer to the wall. Note that this effect is not apparent from the presented plots as we chose to present the frequency in its ``Taylor representation", i.e. divided by the mean velocity.

A second important observation is that the frequency distributions are not sharp, but display a considerable Doppler broadening, i.e. a range of frequencies contributes to the spectral energy content of a single wavenumber. The majority of this effect can be accounted for by random sweeping effects of the smaller scales by the large scales. The broadening, which is a signature of temporal decorrelation, is further enhanced by the temporal evolution of the small scales. Additionally one has to keep in mind that the projection to the $k_1$-$\omega$-plane leads to aliasing effects which also contribute to the Doppler broadening. The Doppler broadening decreases with increasing distance from the wall because the variance of turbulent fluctuations (which are dominated by large-scale contributions) decreases. 

Spanwise wavenumber-frequency spectra are shown in the left panels of figure \ref{fig:span_wf_spec_vs_model}. As the mean velocity has a streamwise component only, the characteristic tilt of the streamwise wavenumber-frequency spectrum is absent in this projection. Still, the mean velocity advection influences the frequency scatter in this plot; the unprojected spectrum $E_{11}(\bs k,\omega;z)$ can be pictured as an ellipsoid with a tilted orientation in the coordinate system spanned by $k_1$, $k_2$, and $\omega$. The central axis of this ellipsoid is in good approximation given by the relation $\omega = \bs k \cdot \bs U$, where $\bs U$ is the mean velocity carrying the velocity fluctuations. When projected to the $k_2$-$\omega$-plane, the Doppler shift contributes to the broad frequency distribution observed from the LES data. Of course, large-scale random-sweeping effects further increase the broadening.

\section{Random-sweeping model for the wavenumber-frequency spectrum}

\begin{figure}
\begin{center}
    \includegraphics[width=0.58\textwidth]{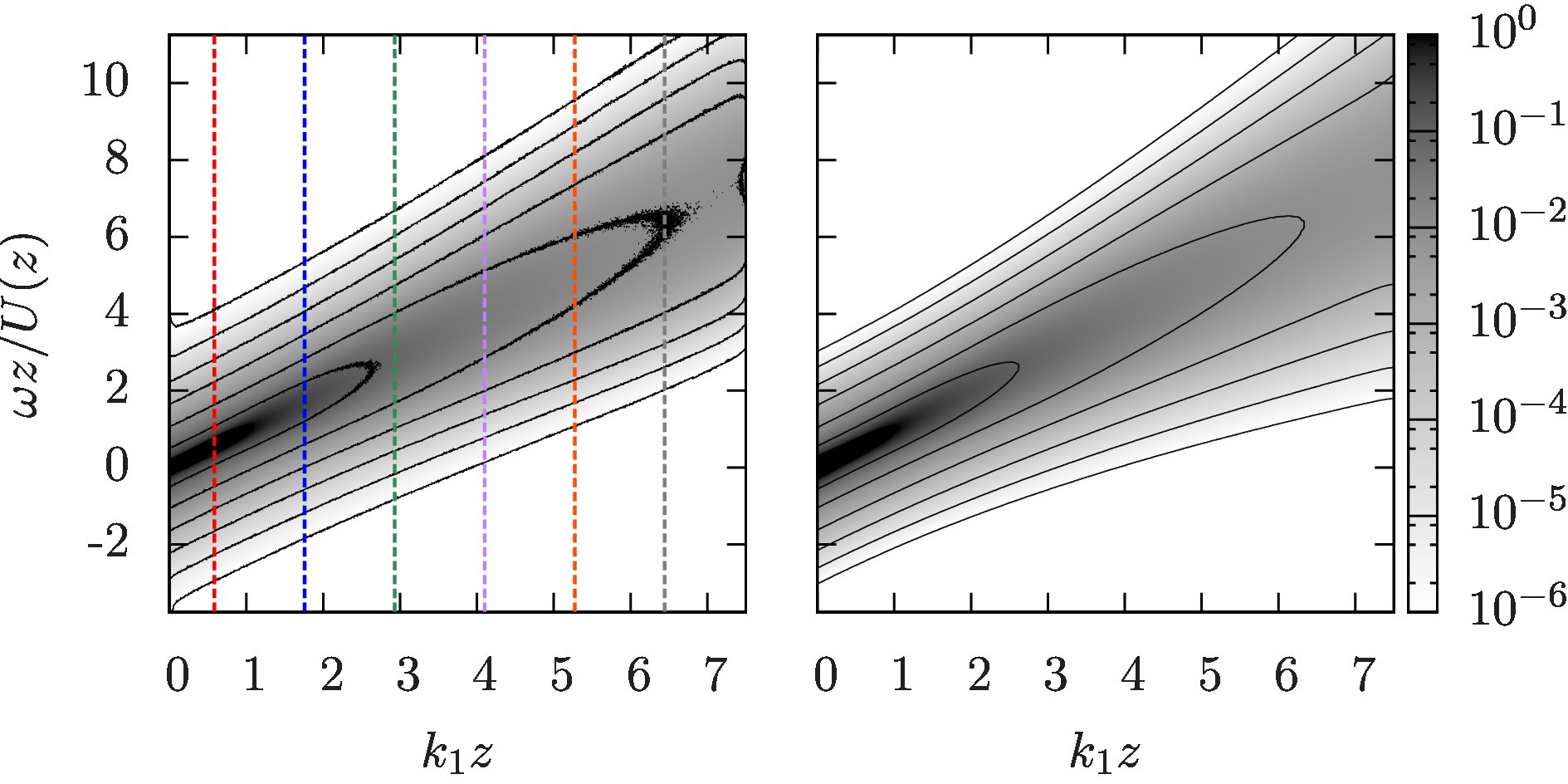}
    \includegraphics[width=0.41\textwidth]{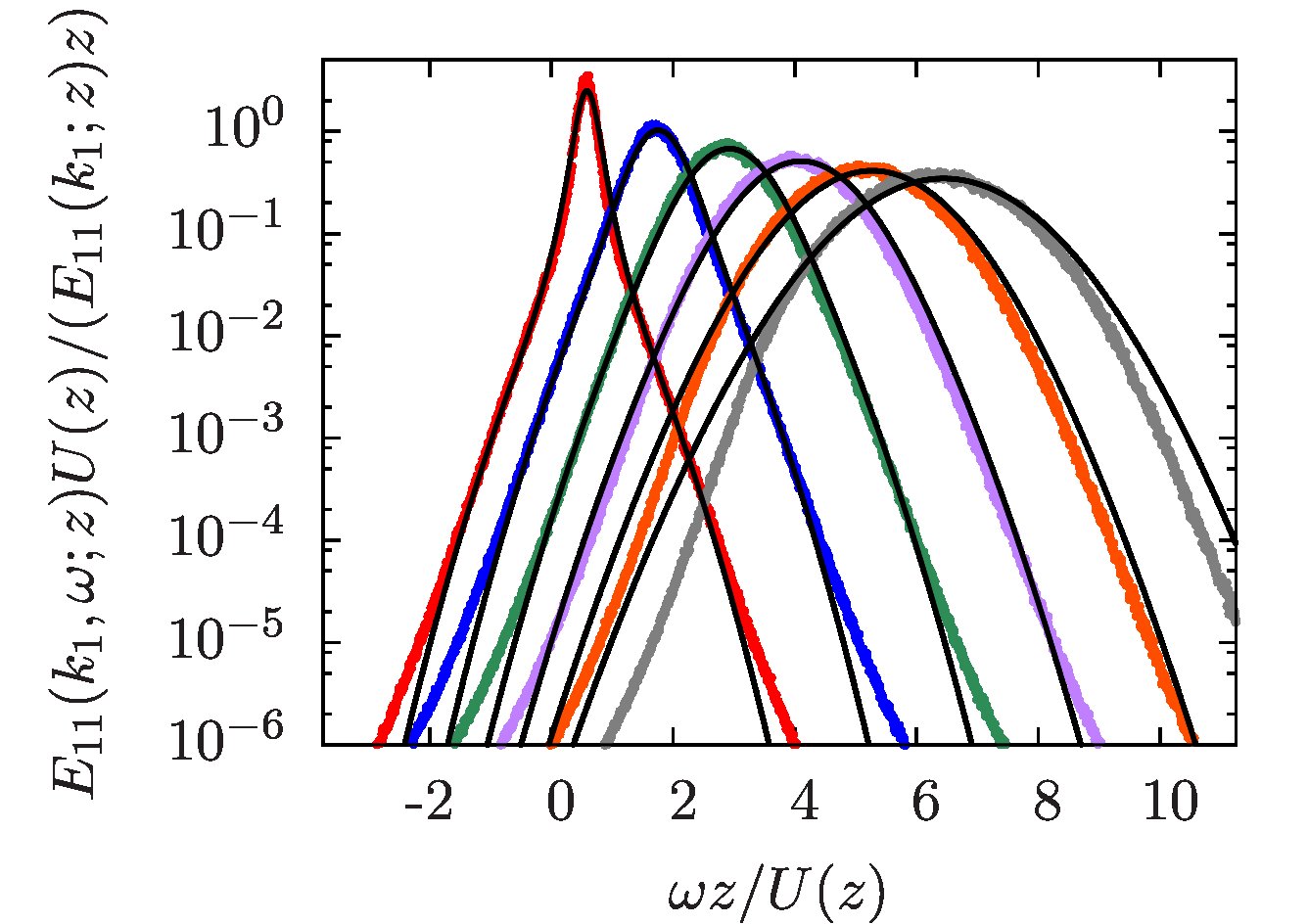}
    \includegraphics[width=0.58\textwidth]{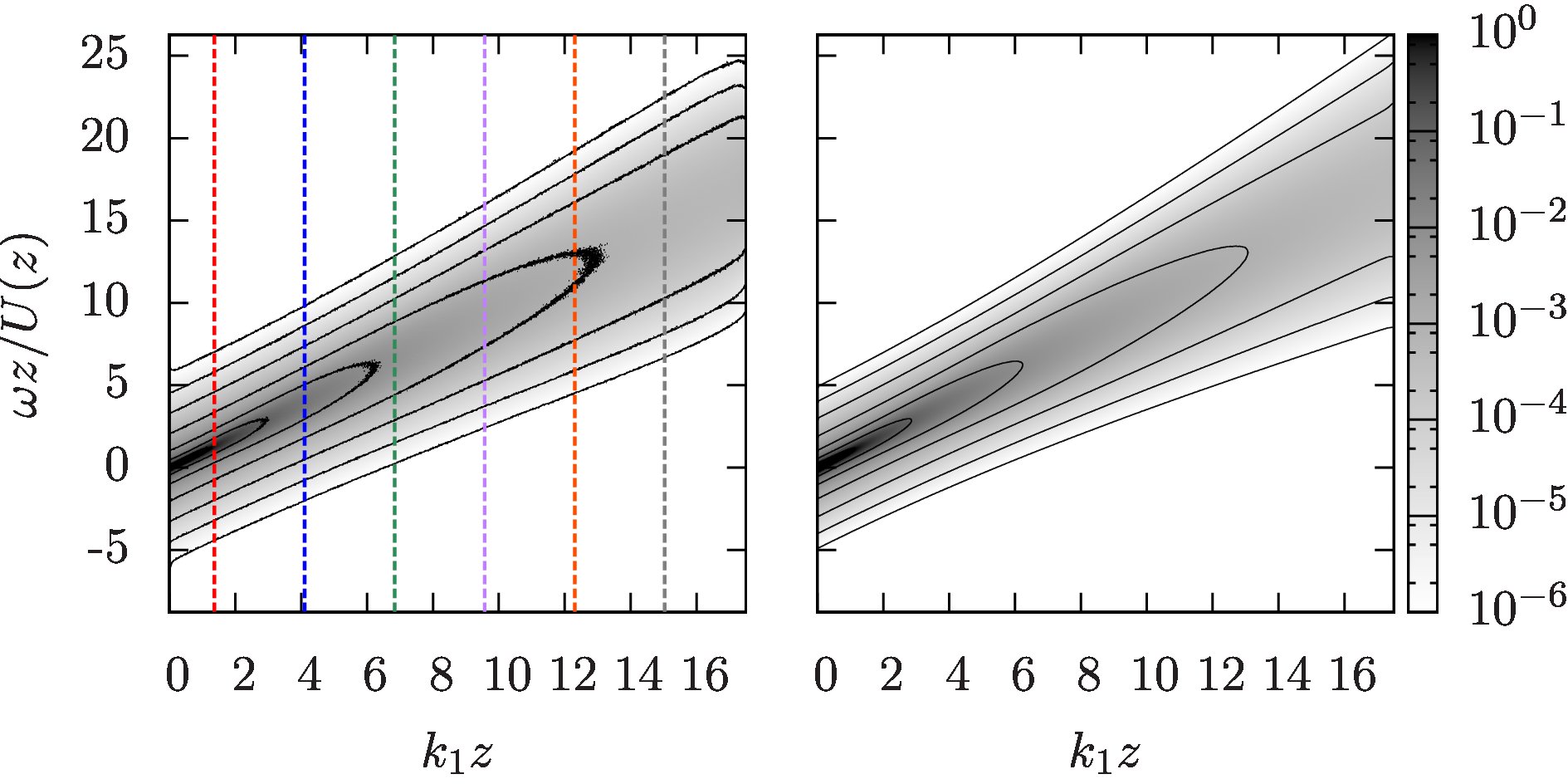}
    \includegraphics[width=0.41\textwidth]{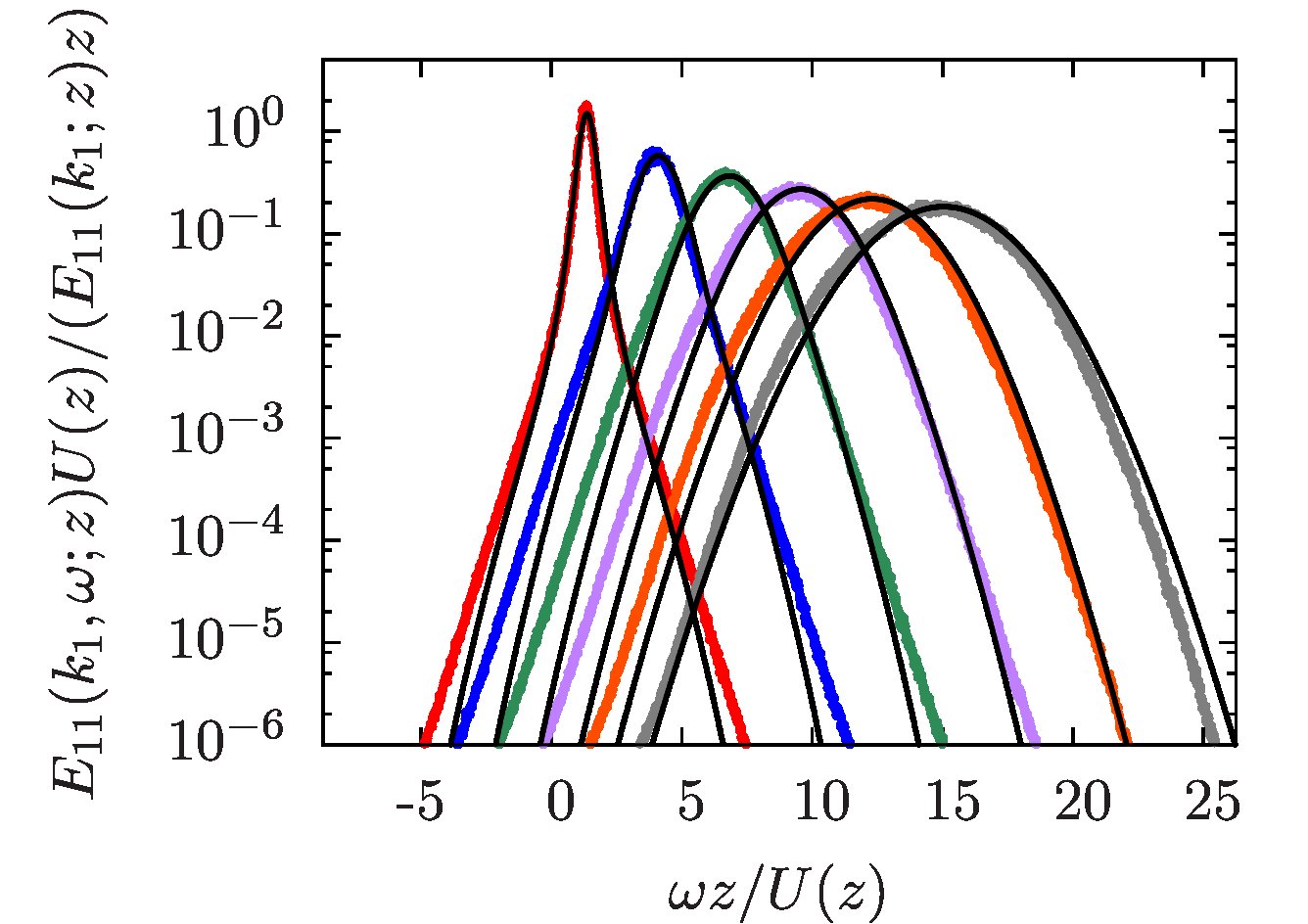}
    \includegraphics[width=0.58\textwidth]{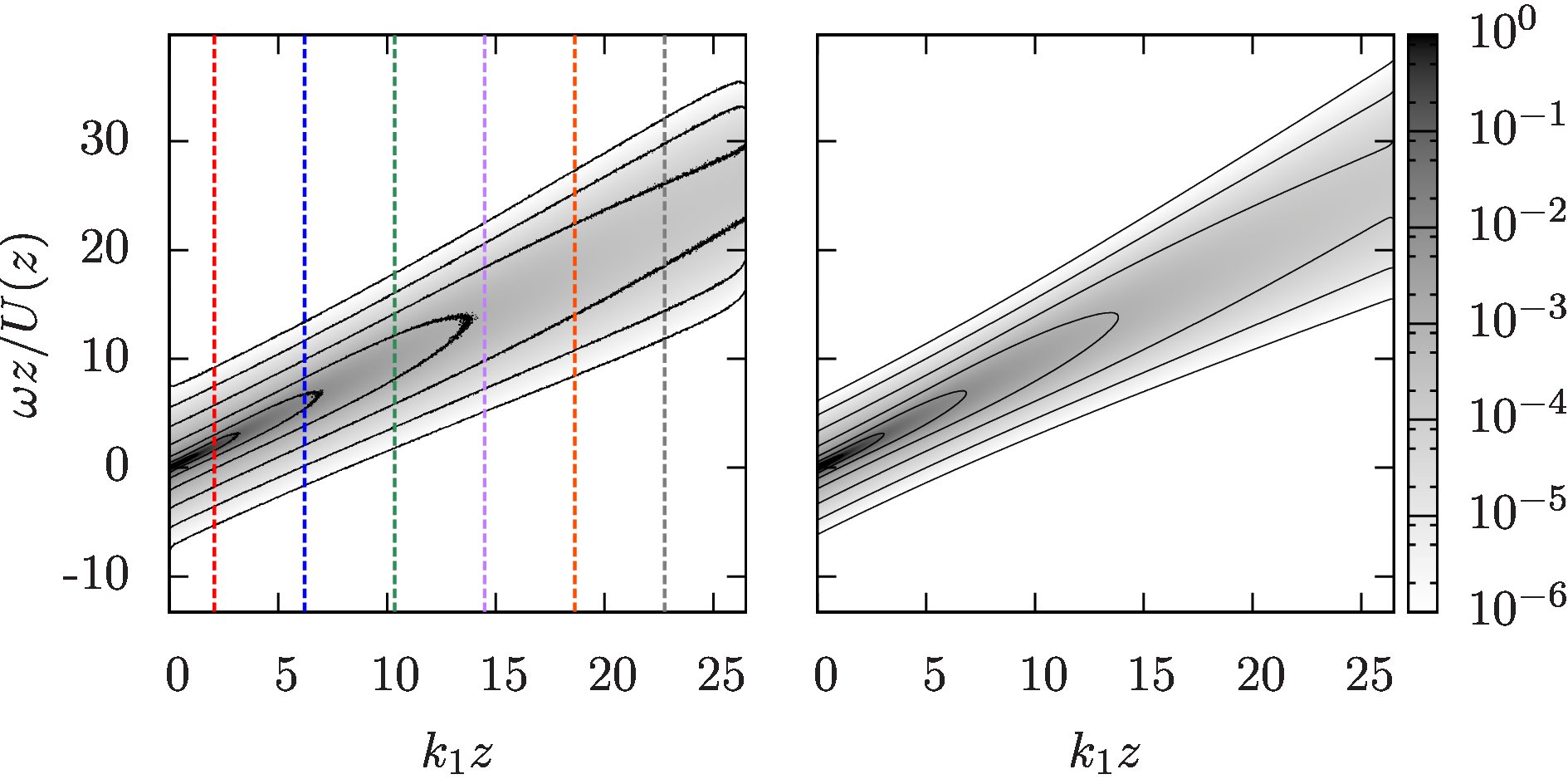}
    \includegraphics[width=0.41\textwidth]{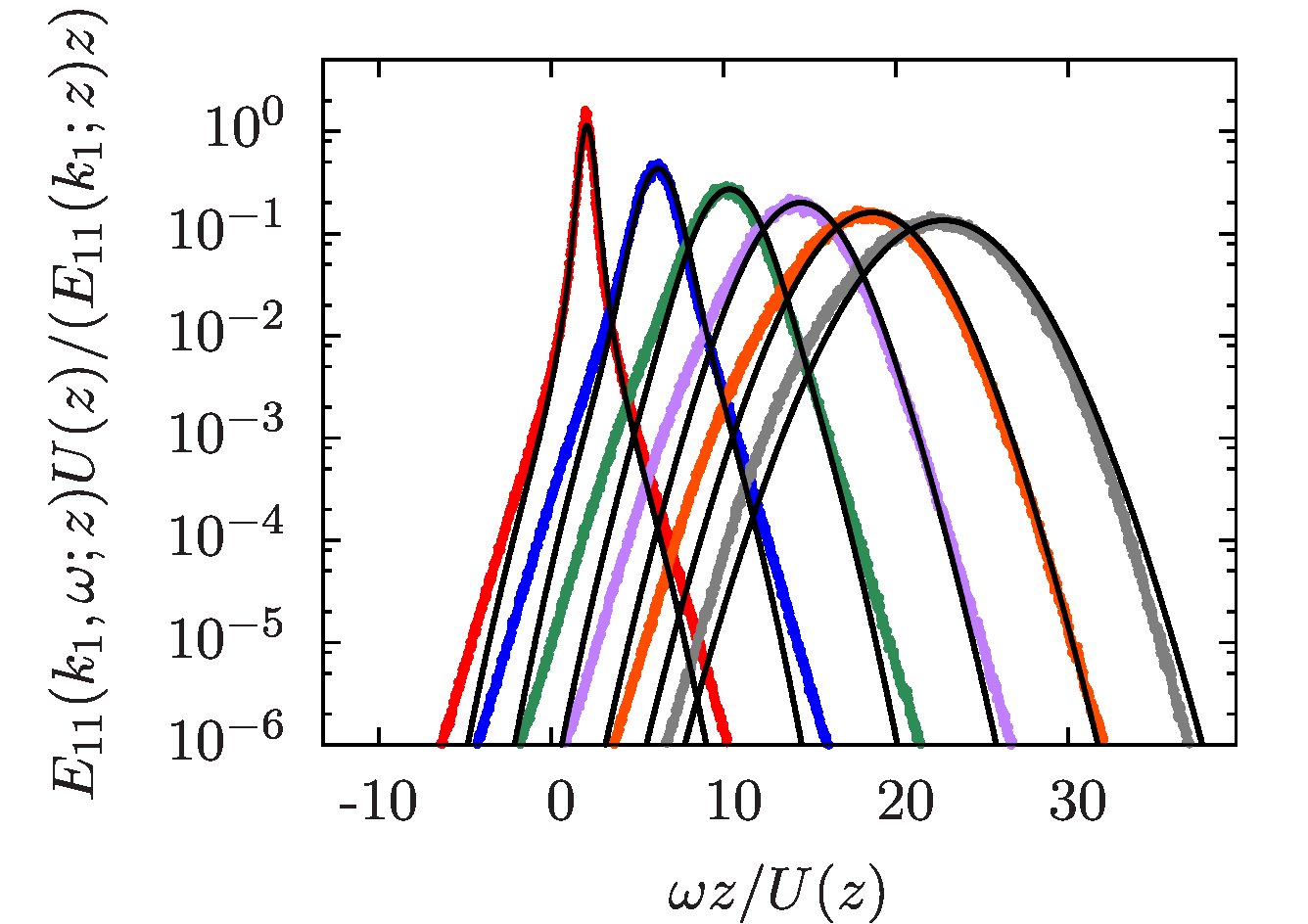}
    \includegraphics[width=0.58\textwidth]{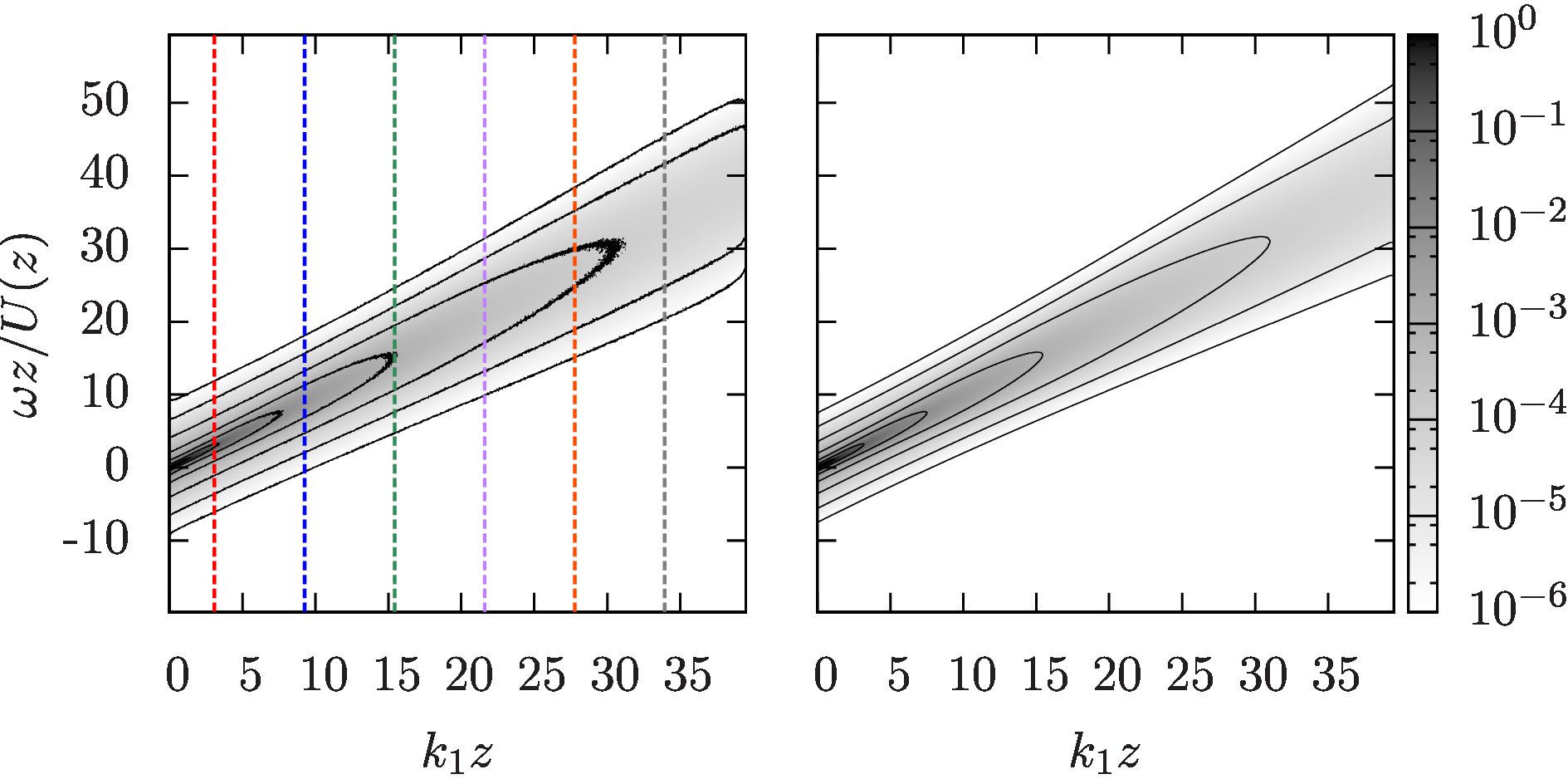}
    \includegraphics[width=0.41\textwidth]{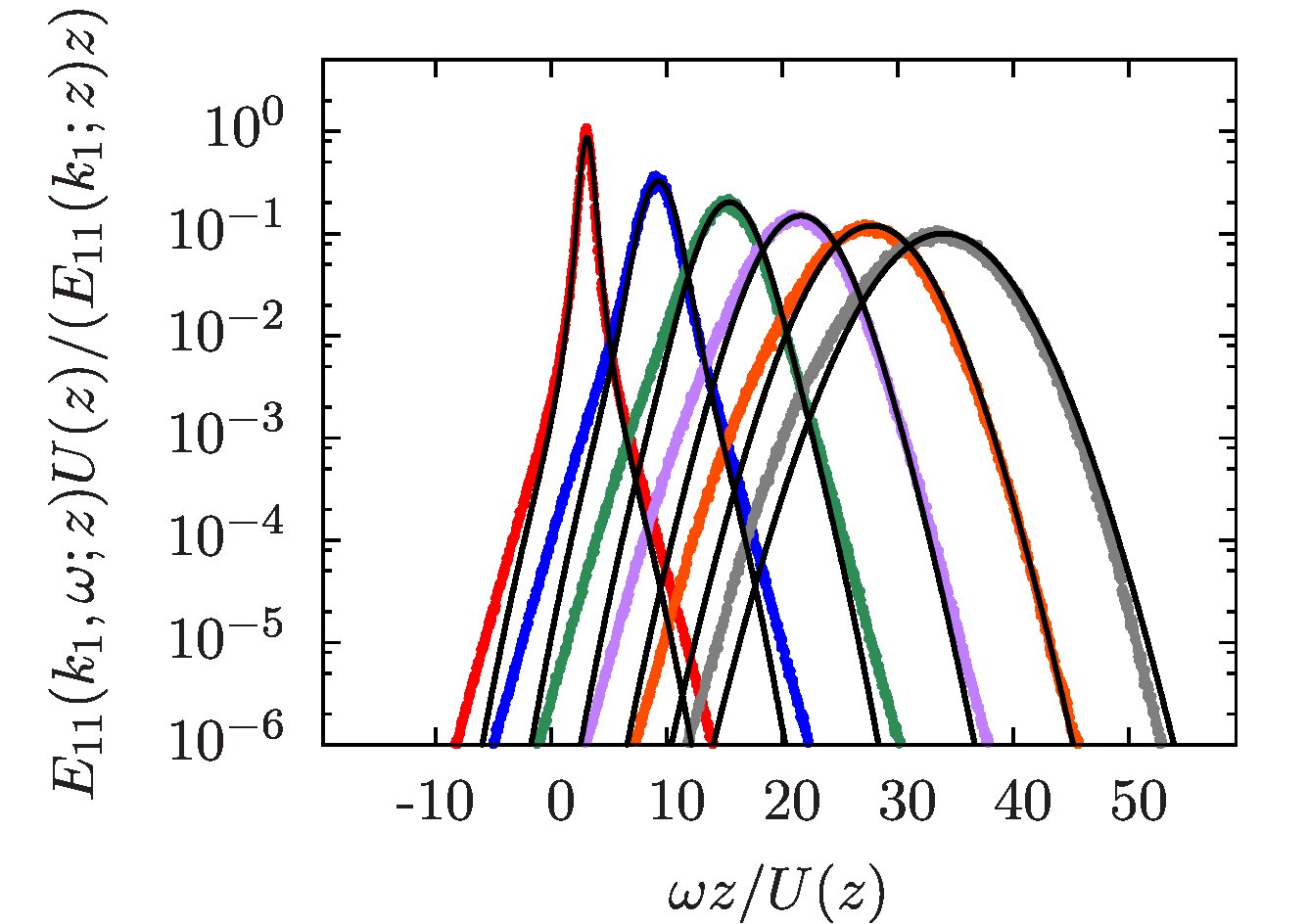}
    \includegraphics[width=0.58\textwidth]{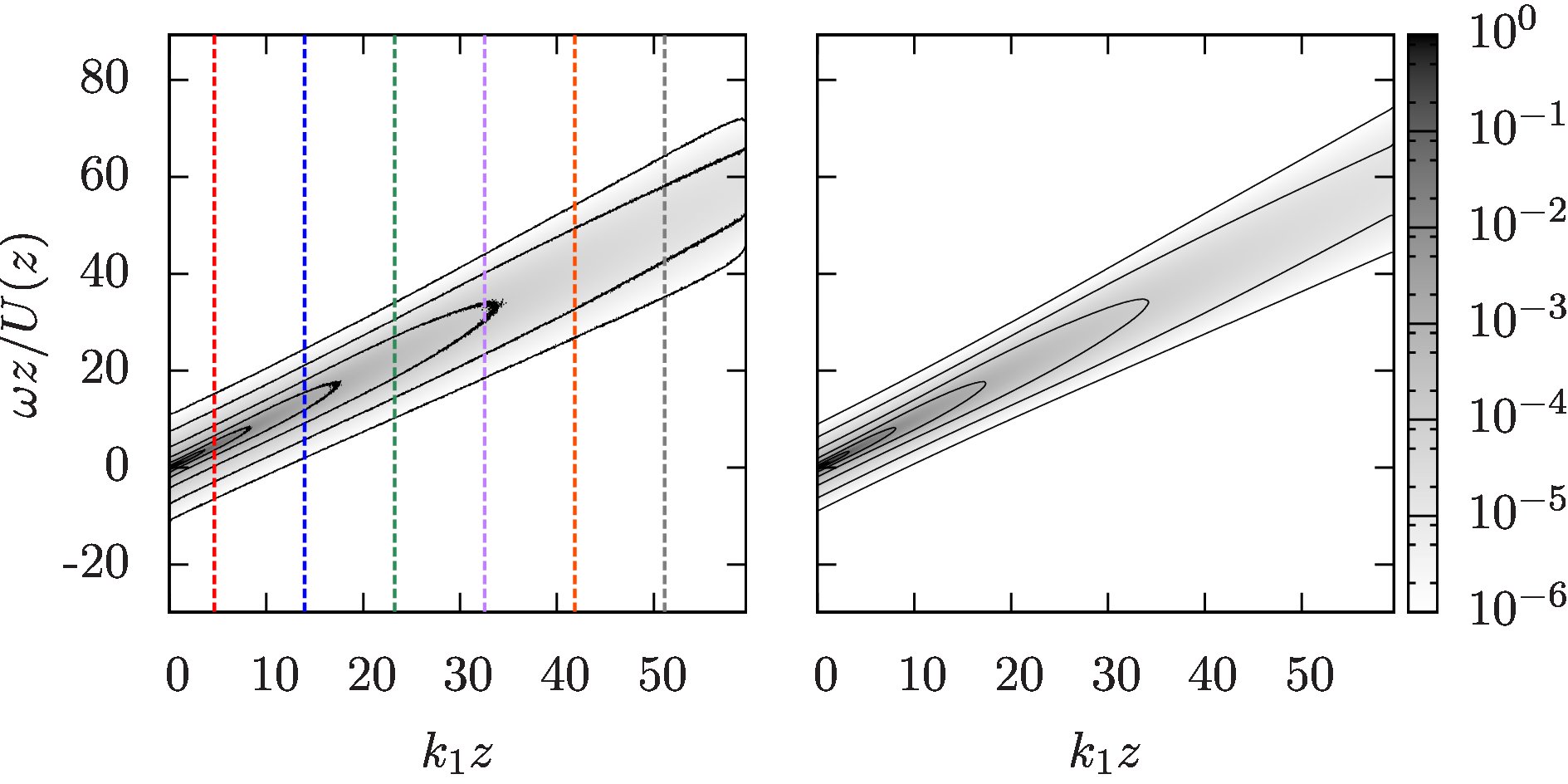}
    \includegraphics[width=0.41\textwidth]{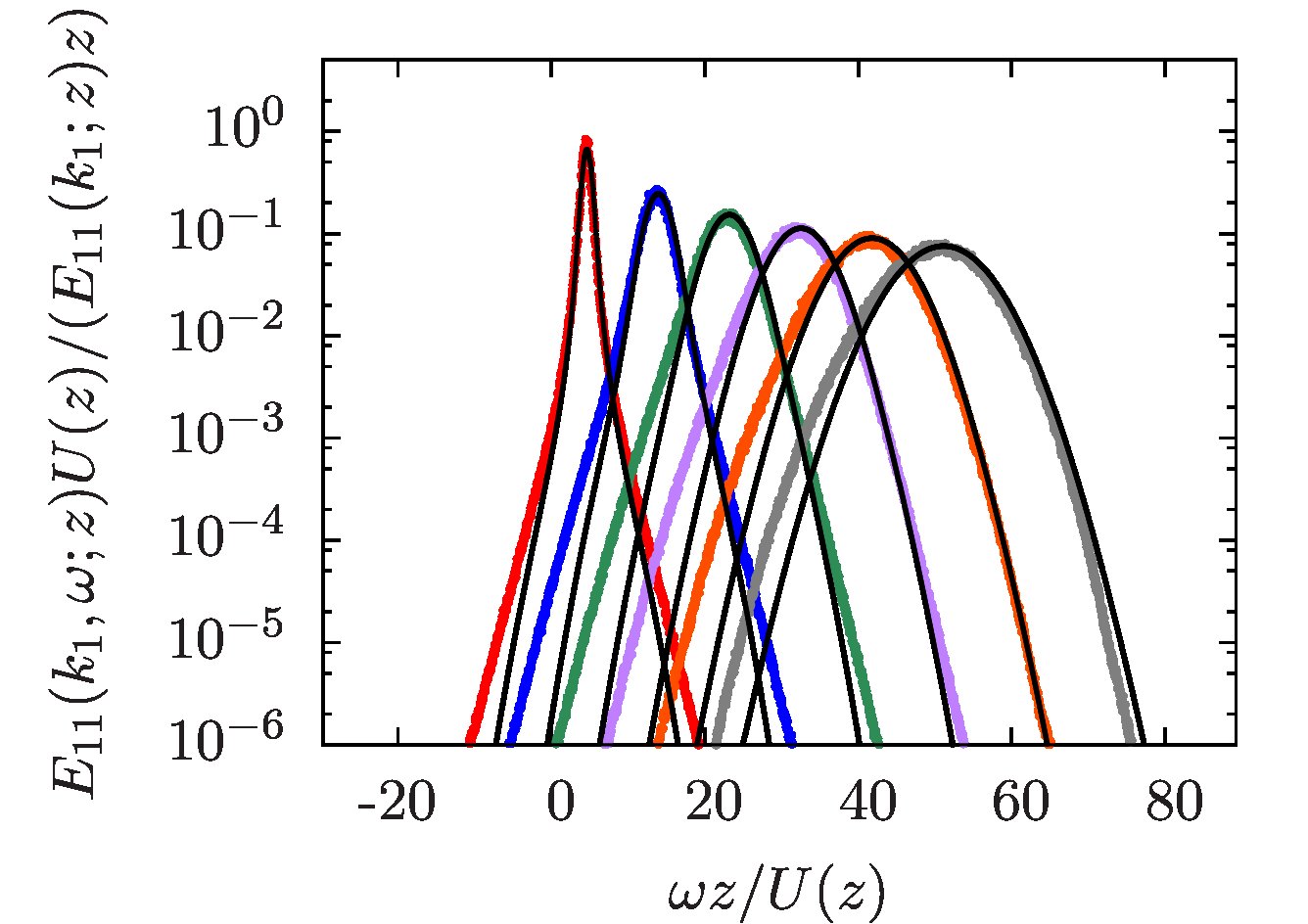}
\end{center}
\caption{Streamwise wavenumber-frequency spectrum for the streamwise velocity component. From top to bottom $z/H \in \lbrace 0.0293,0.0684,0.104,0.154,0.232 \rbrace$. Left: LES data, middle: prediction based on random sweeping hypothesis, right: cuts comparing LES data (colored dots) and random sweeping model (black lines). The axis ranges are chosen such that all plots show the same range in $k_1$ and $\omega/U(z)$.}
\label{fig:stream_wf_spec_vs_model}
\end{figure}

\begin{figure}
\begin{center}
    \includegraphics[width=0.58\textwidth]{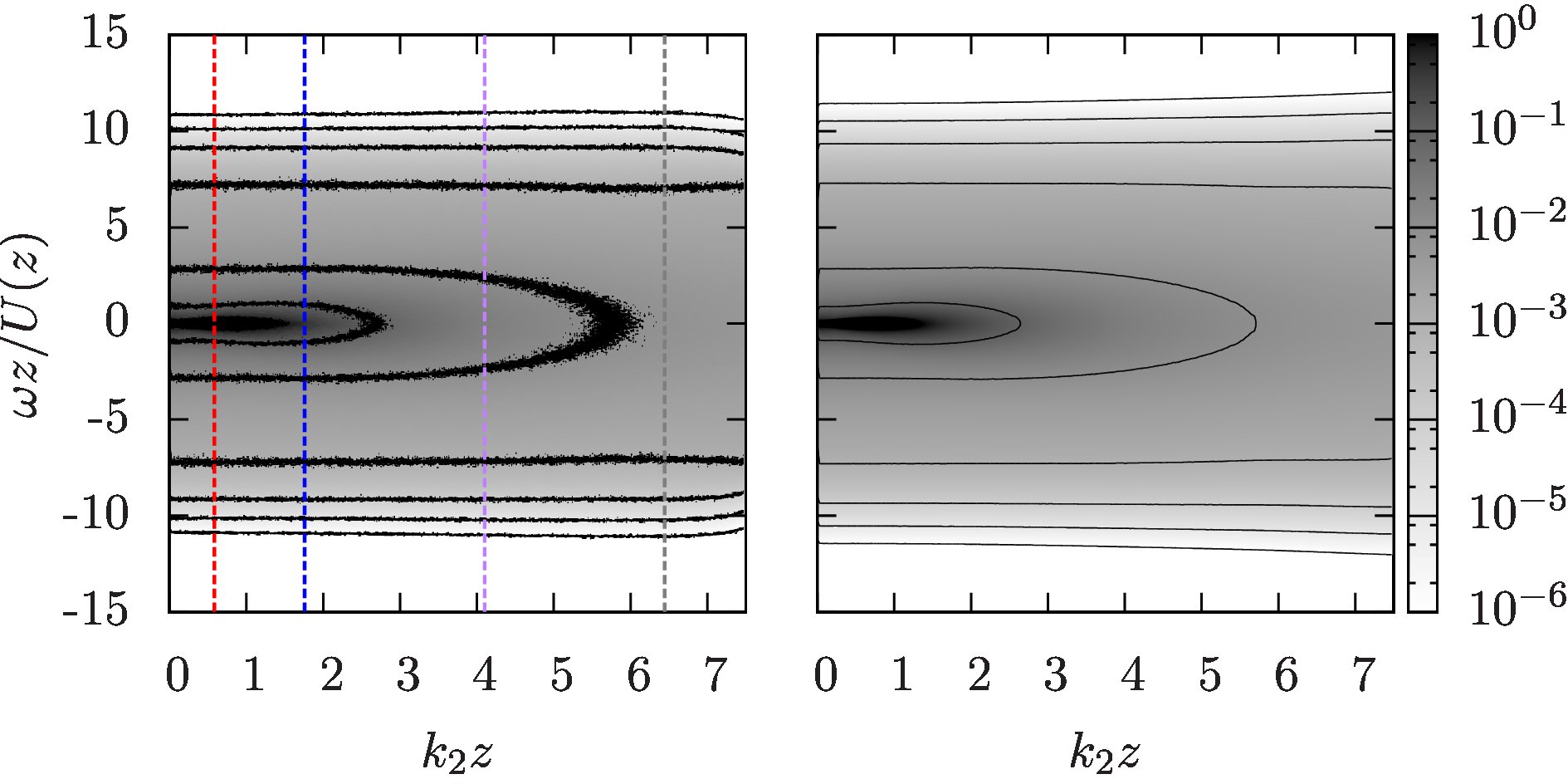}
    \includegraphics[width=0.41\textwidth]{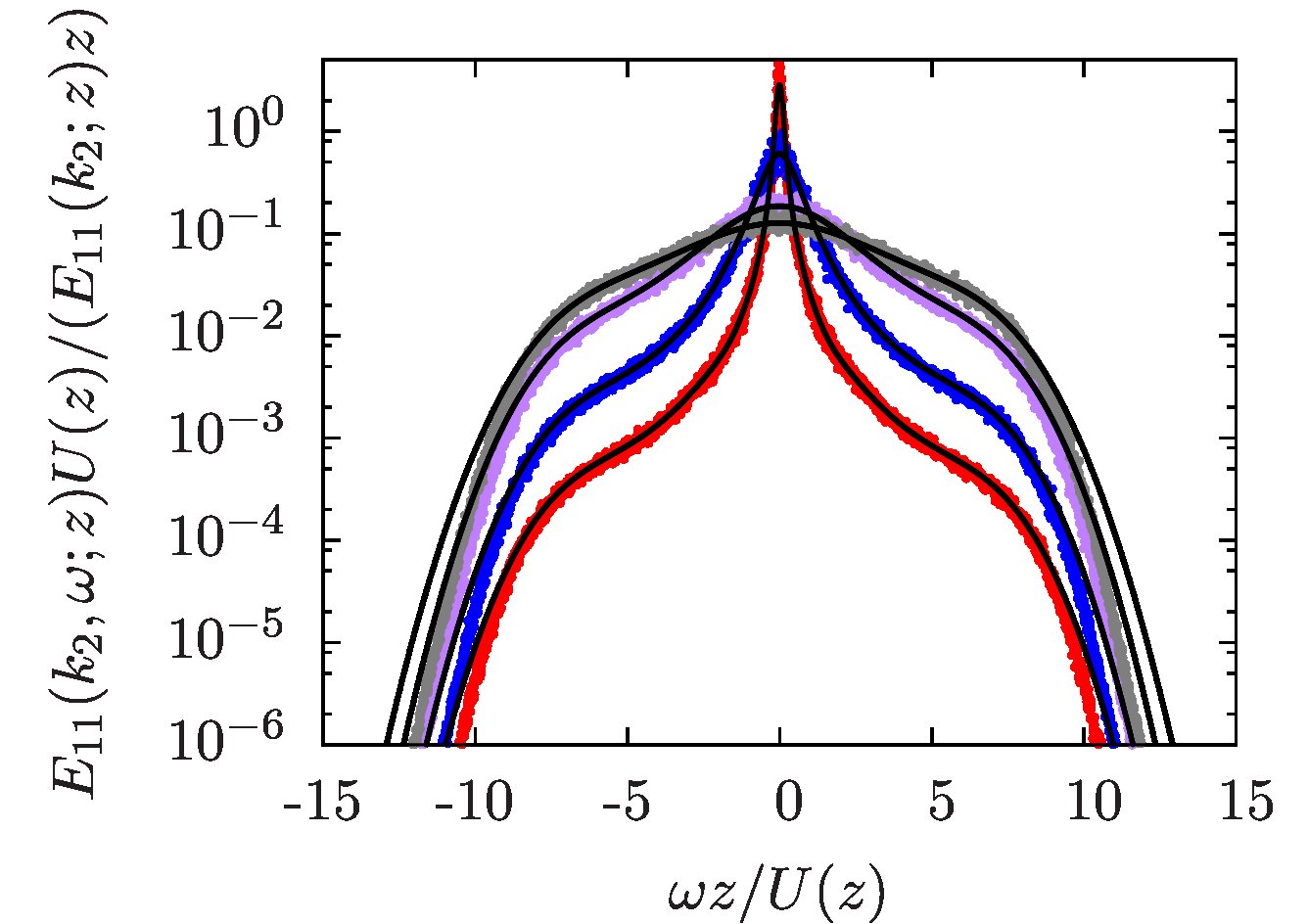}
    \includegraphics[width=0.58\textwidth]{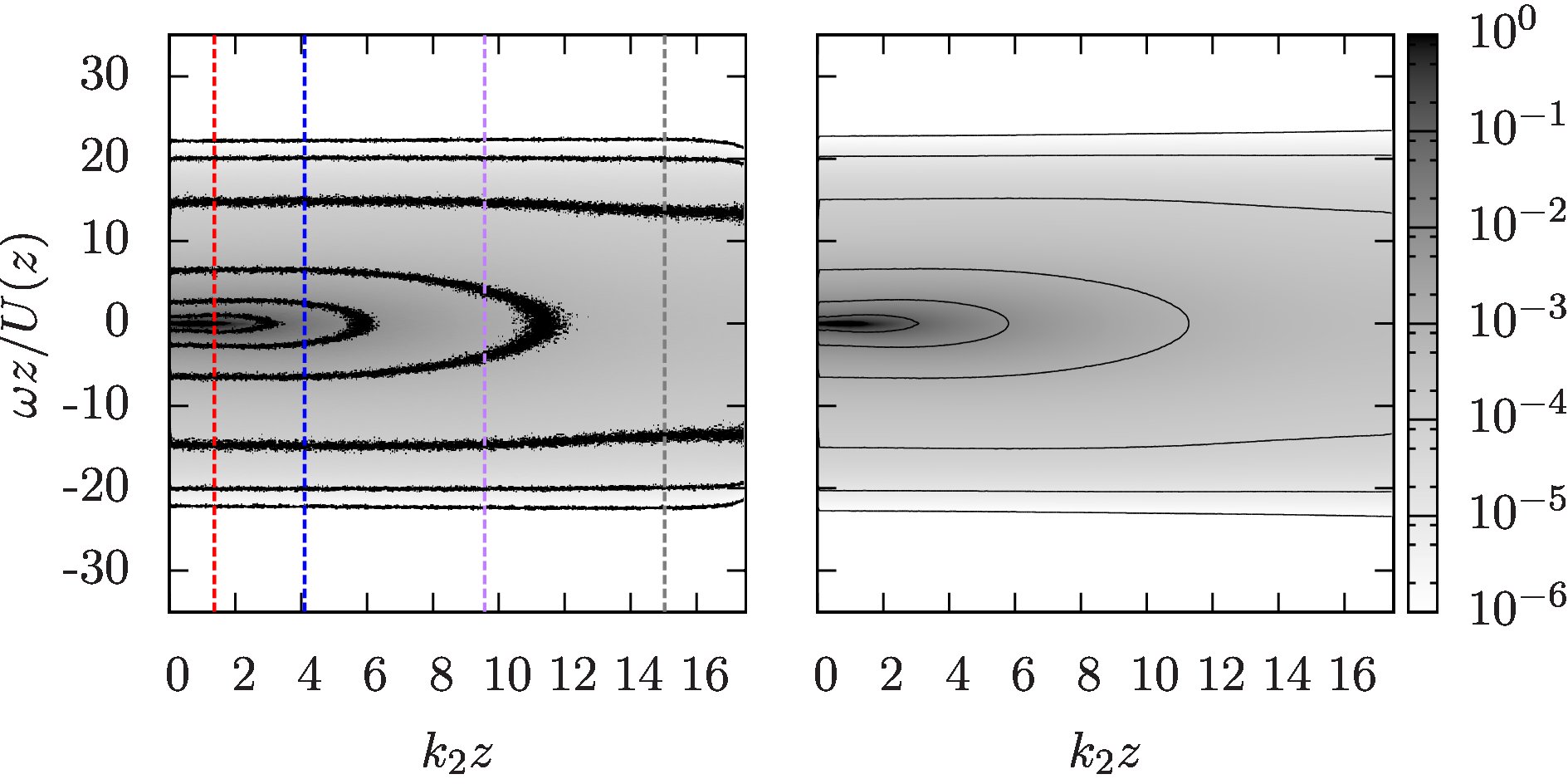}
    \includegraphics[width=0.41\textwidth]{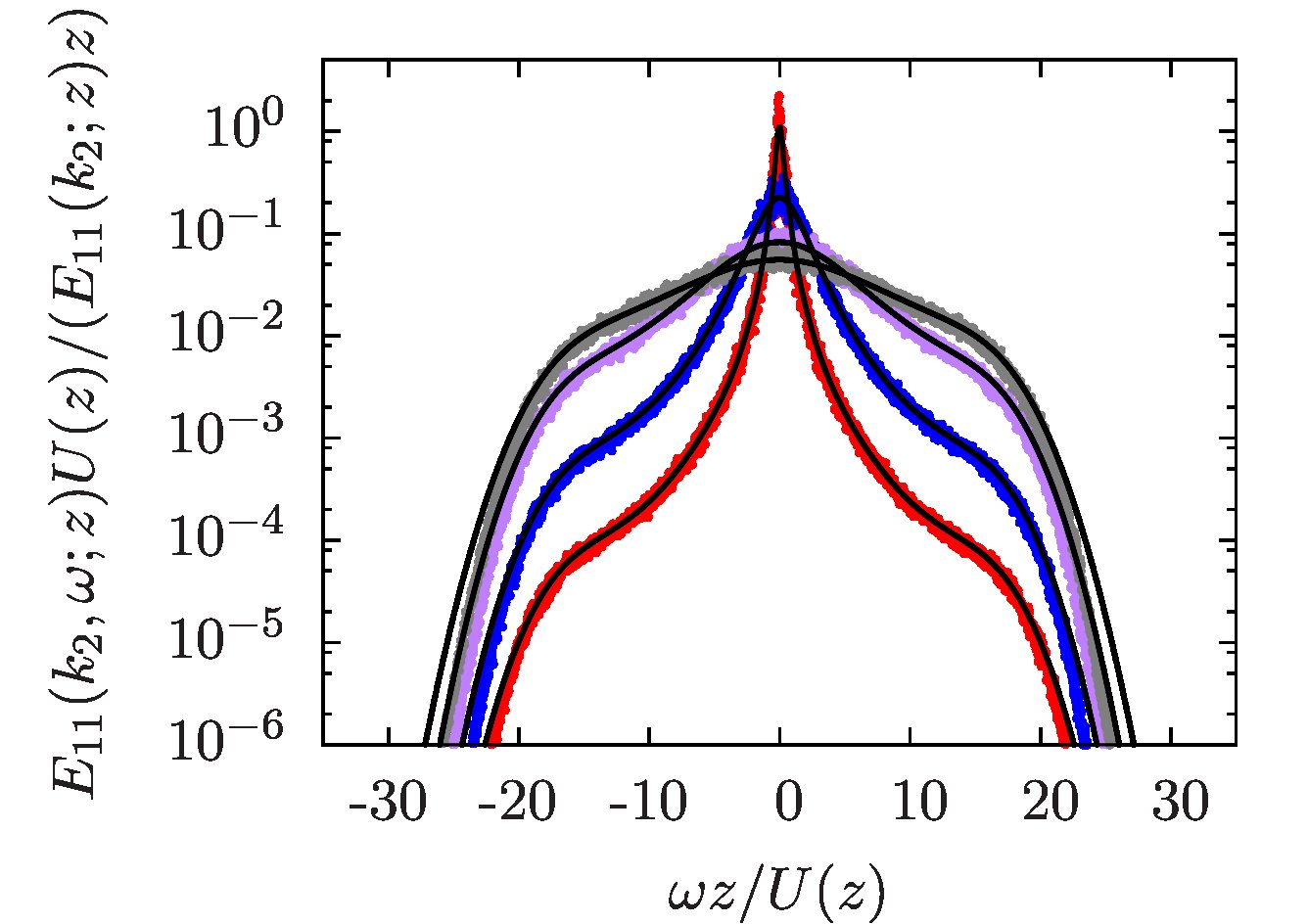}
    \includegraphics[width=0.58\textwidth]{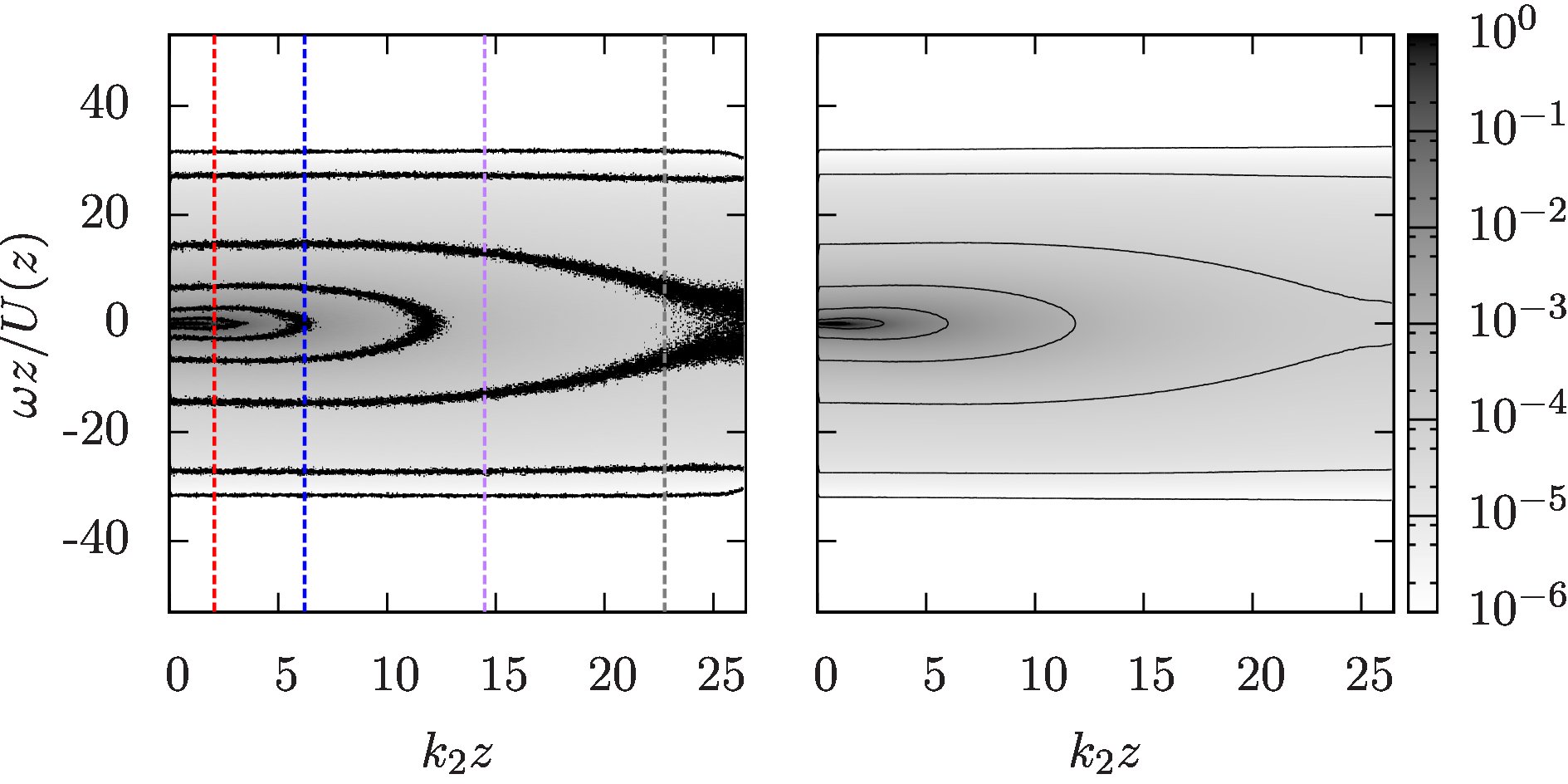}
    \includegraphics[width=0.41\textwidth]{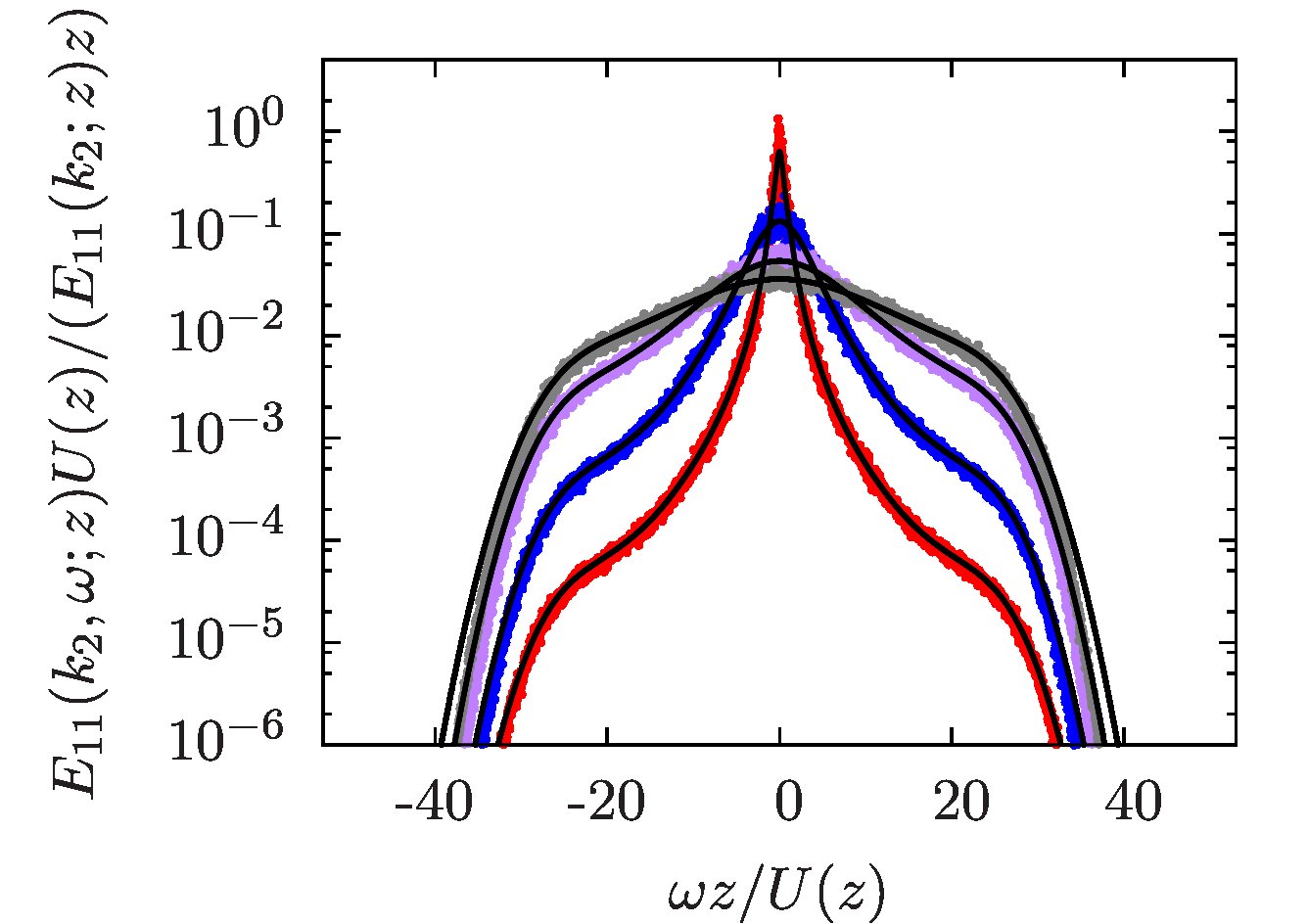}
    \includegraphics[width=0.58\textwidth]{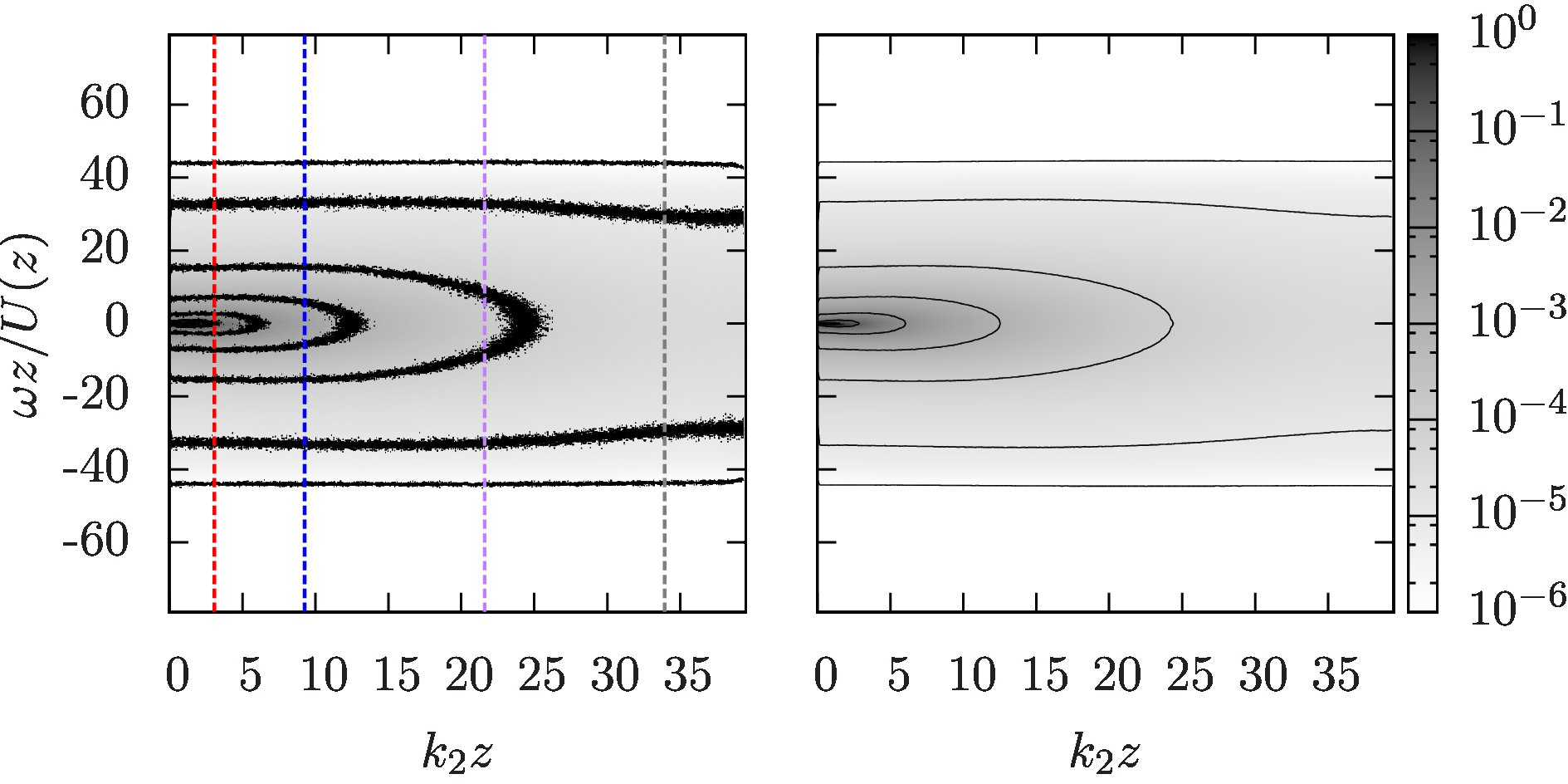}
    \includegraphics[width=0.41\textwidth]{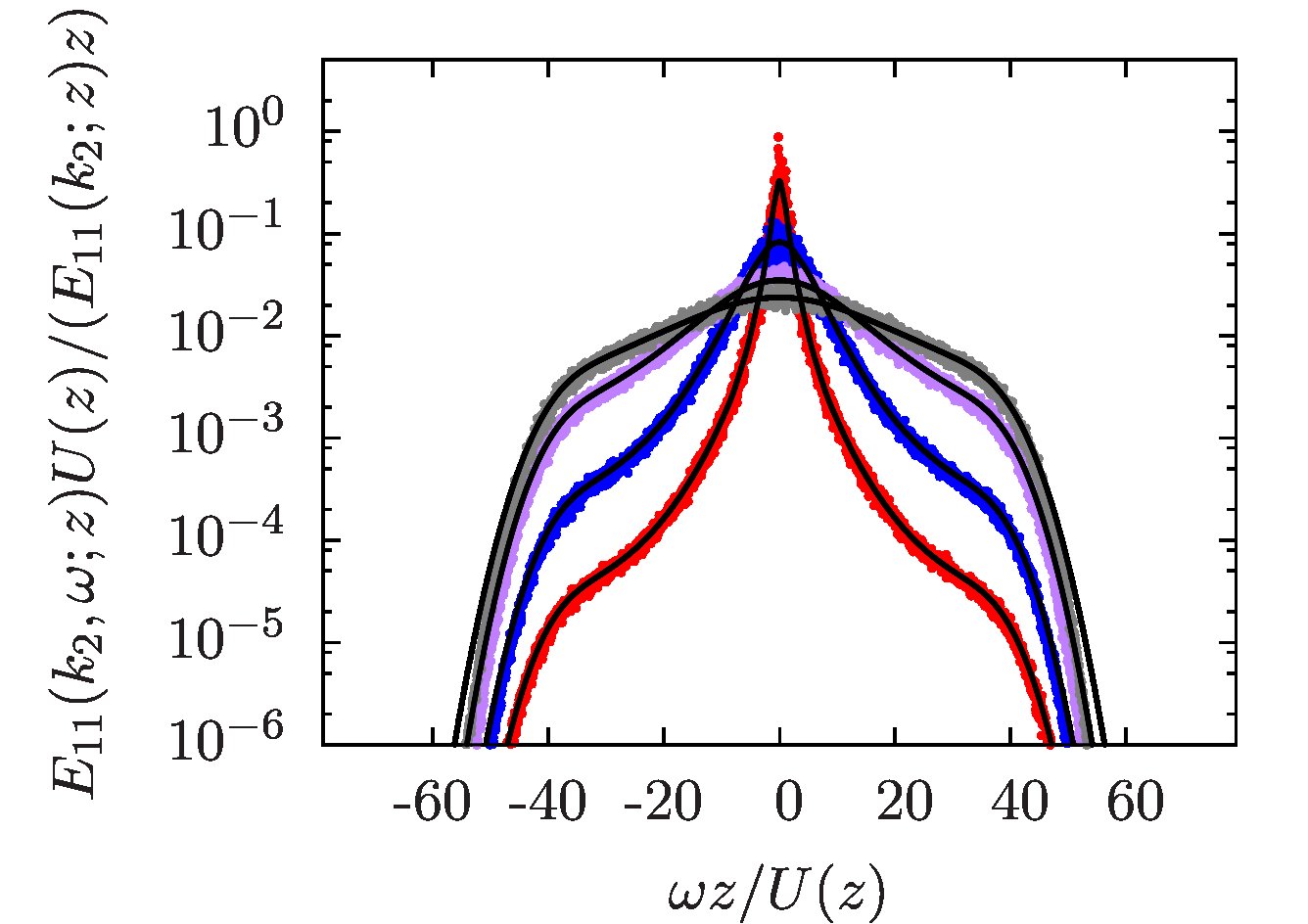}
    \includegraphics[width=0.58\textwidth]{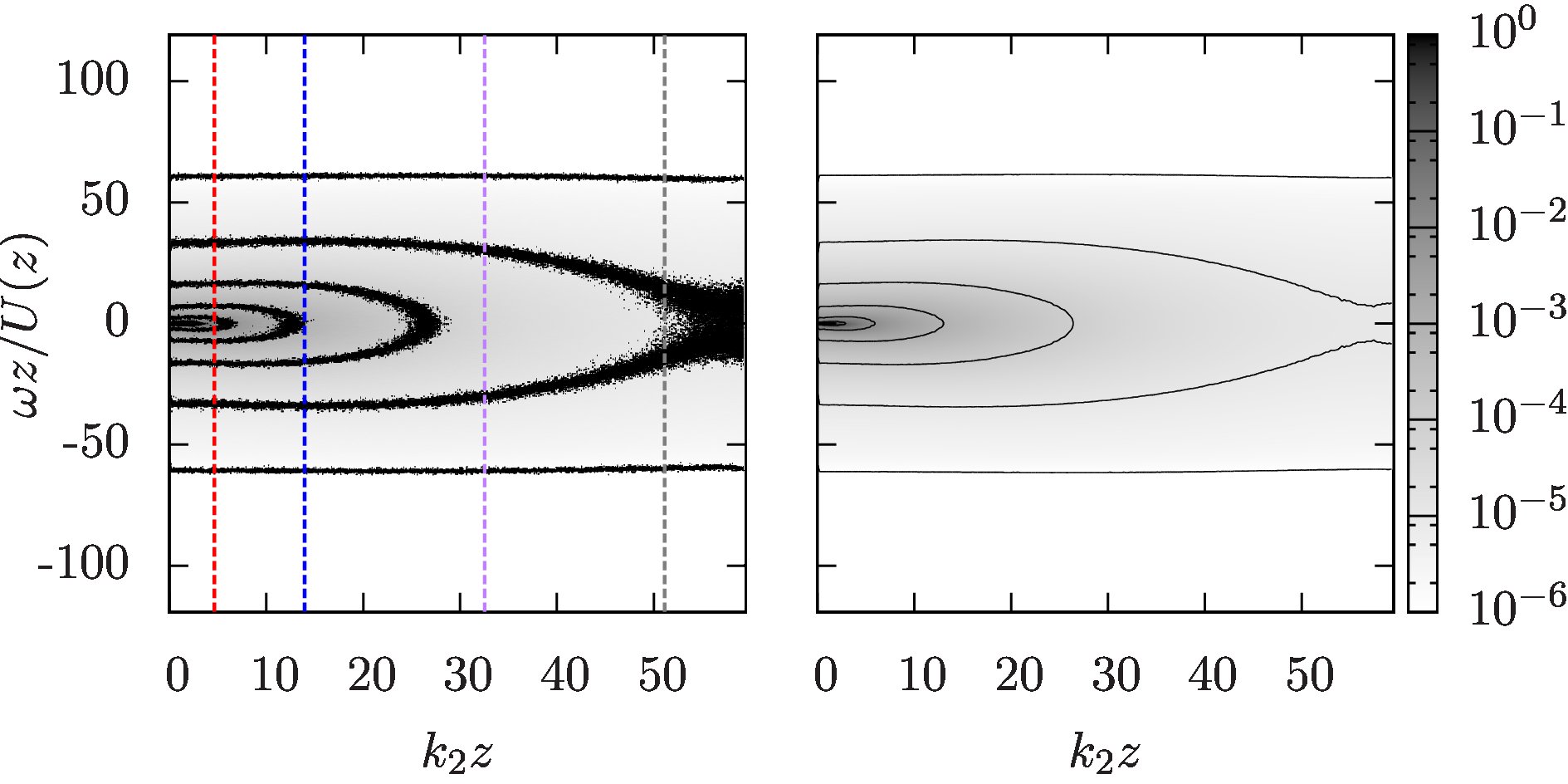}
    \includegraphics[width=0.41\textwidth]{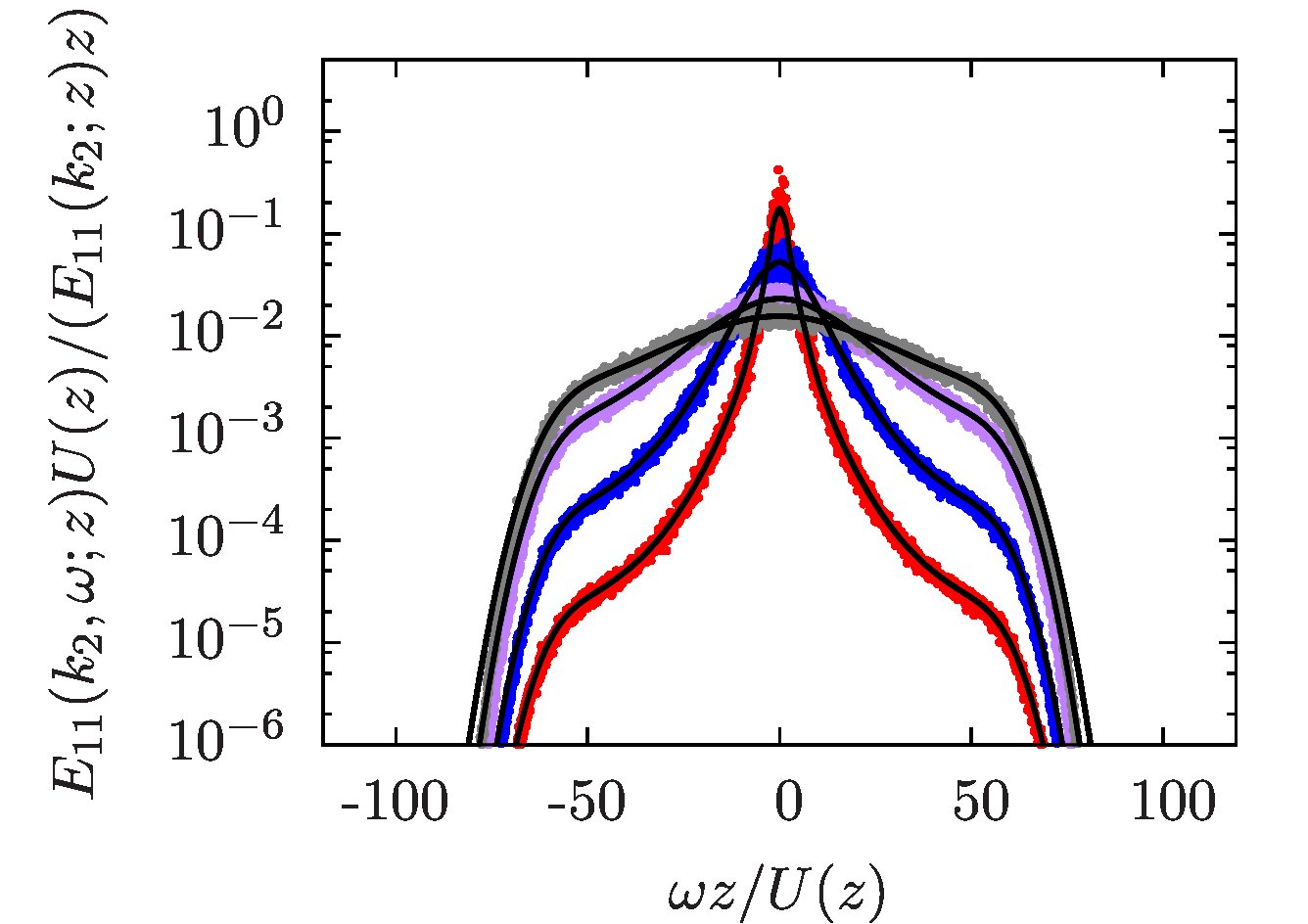}
\end{center}
\caption{Spanwise wavenumber-frequency spectrum for the streamwise velocity component. From top to bottom $z/H \in \lbrace 0.0293,0.0684,0.104,0.154,0.232 \rbrace$. Left: LES data, middle: prediction based on random sweeping hypothesis, right: cuts comparing LES data (colored dots) and random sweeping model (black lines). The axis ranges are chosen such that all plots show the same range in $k_2$ and $\omega/U(z)$.}
\label{fig:span_wf_spec_vs_model}
\end{figure}

The main features of the wavenumber-frequency spectra obtained from LES can be captured in a simple model based on the Kraichnan-Tennekes random-sweeping hypothesis \cite{kraichnan64pof,tennekes75jfm} with mean flow. The model has been discussed for wall-bounded flows in \cite{wilczek15jfm}, to which we refer the reader for more details and further references.
It is centered around the idea that smaller-scale streamwise velocity fluctuations $u$ are carried with a mean velocity $\bs U = U \bs e_1$ and a large-scale random-sweeping velocity $\bs v$. We restrict ourselves to spectra of $u$ in horizontal planes, and we furthermore consider only a planar random-sweeping velocity with zero mean. Vertical interactions are consequently not taken into account. We also neglect the influence of shear. Additionally, we assume a large scale separation between the velocity fluctuations and the random-sweeping velocity, such that, in lowest-order approximation, the small scales are not distorted by the large scales. To include randomness, the sweeping velocity $\bs v$ is taken constant in space and time, but with a Gaussian ensemble distribution of prescribed covariance.
Based on these assumptions we arrive at the prediction of the joint wavenumber-frequency spectrum of the streamwise velocity component as a product of the wavenumber spectrum and a frequency distribution \cite{wilczek12pre,wilczek15jfm},
\begin{equation}\label{eq:wavenumberfrequencyspectrum}
  E_{11}(\bs k,\omega;z) = E_{11}(\bs k;z) \left[2\pi \left\langle (\bs v \cdot \bs k)^2  \right\rangle \right]^{-1/2} \exp\left[ -\frac{(\omega - \bs k \cdot \bs U)^2}{2 \left\langle (\bs v \cdot \bs k)^2  \right\rangle}  \right] \, .
\end{equation}
The frequency distribution in this model is Gaussian (which is a direct consequence of the assumption of a large-scale velocity that is constant in time) with wavenumber-dependent Doppler shift induced by the mean flow, and Doppler broadening related to the random-sweeping velocity.

To test the prediction for the wavenumber-frequency spectrum based on the random-sweeping hypothesis, we take the spectra from LES presented in the preceding section as a benchmark and compare them to the spectra given by the right-hand side of \eqref{eq:wavenumberfrequencyspectrum}. To evaluate the right-hand side of \eqref{eq:wavenumberfrequencyspectrum} we determine the wavenumber spectrum $E_{11}(\bs k;z)$ from the LES data, along with the mean velocity and the covariance of the velocity fluctuations.

The projections to the $k_1$-$\omega$-plane and the $k_2$-$\omega$-plane are shown in figures \ref{fig:stream_wf_spec_vs_model} and \ref{fig:span_wf_spec_vs_model}, respectively. The middle panels of both figures show the right-hand side of \eqref{eq:wavenumberfrequencyspectrum} whereas the right panels show a direct comparison of the frequency distributions of the directly estimated spectra (colored dots) and the random-sweeping prediction (black lines). Given the underlying simplifying assumptions made in the derivation of the model, a remarkable agreement is observed. However, some details differ; most strikingly, the frequency distributions of the $k_1$-$\omega$ spectra from LES show a subtle skewness which is not captured by the random-sweeping model.

\section{Recapitulation of a model spectrum}

\begin{figure}
\begin{center}
    \includegraphics[width=0.8\textwidth]{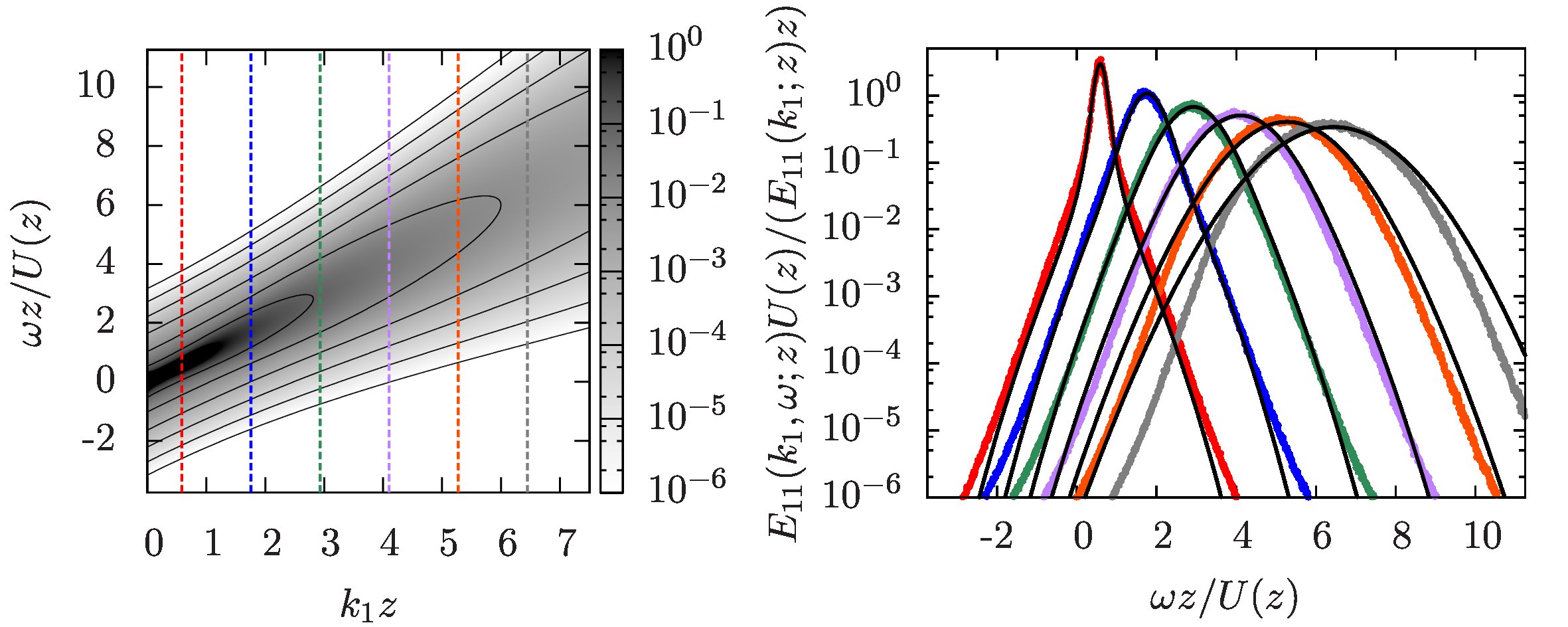}
    \includegraphics[width=0.8\textwidth]{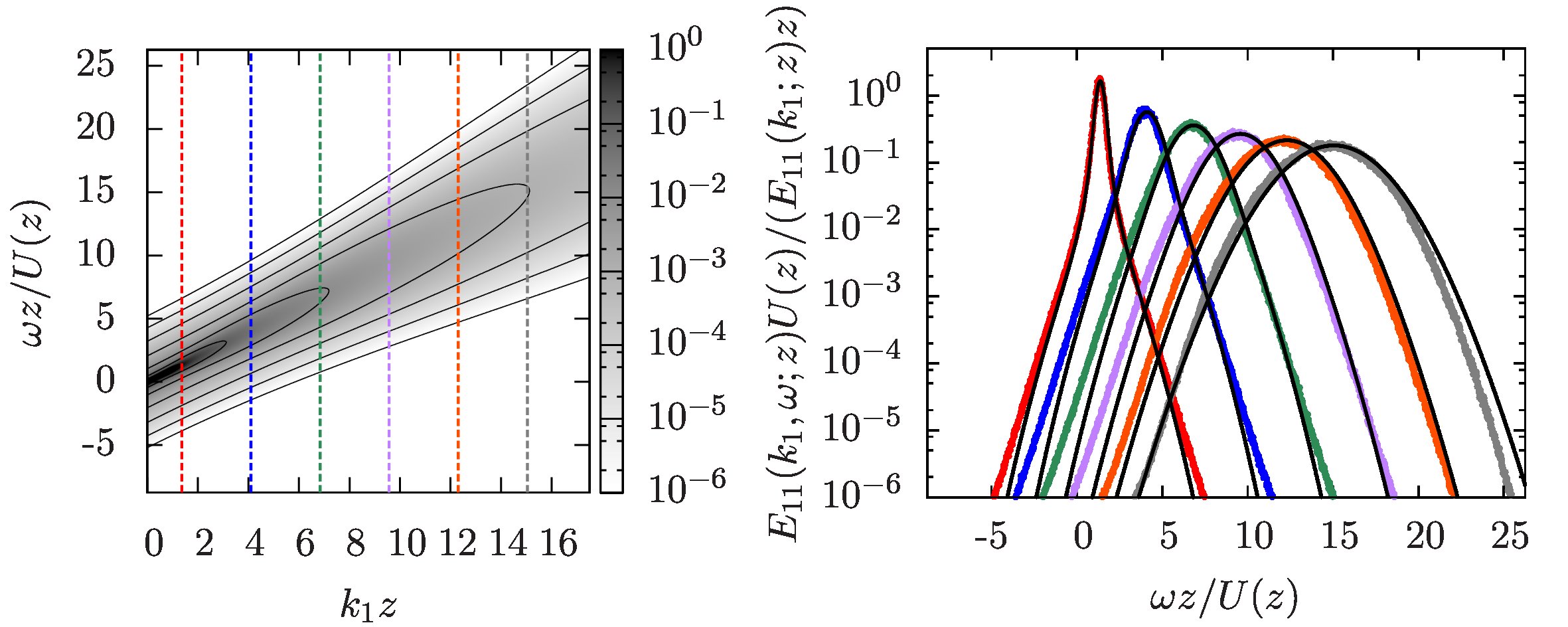}
    \includegraphics[width=0.8\textwidth]{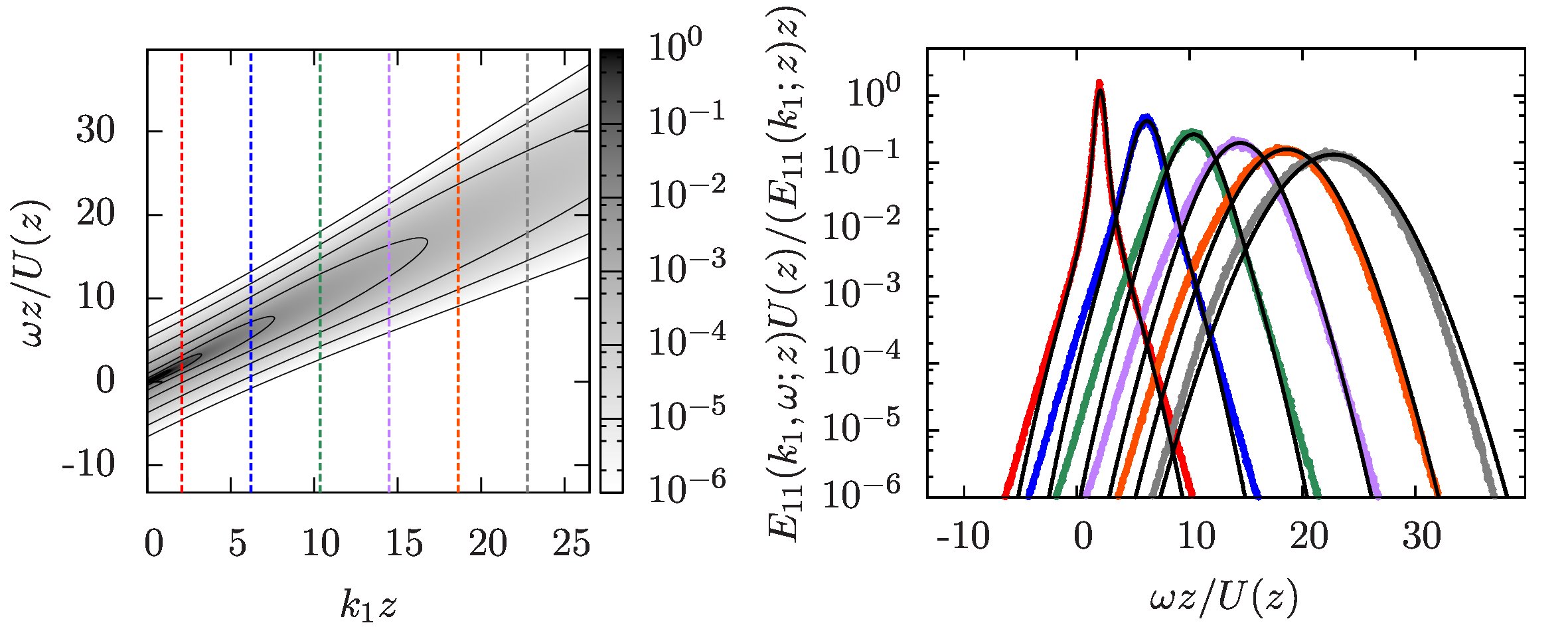}
    \includegraphics[width=0.8\textwidth]{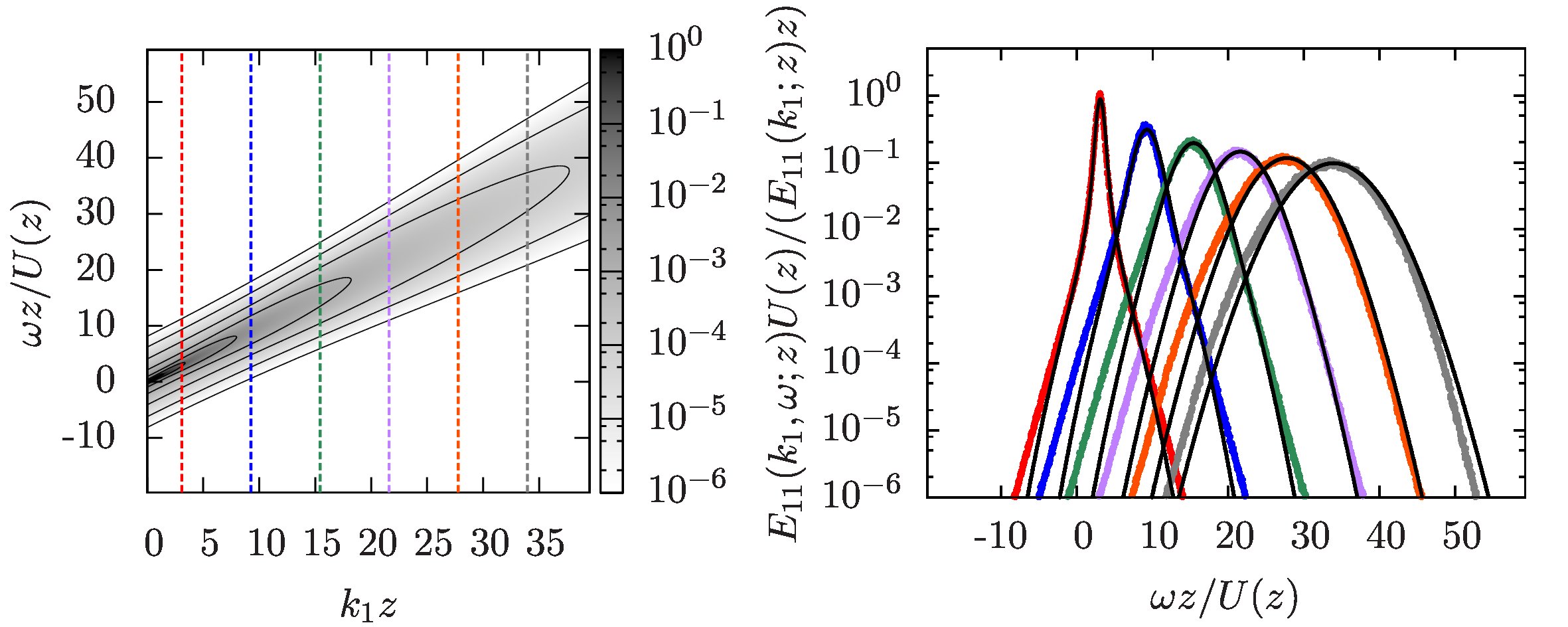}
    \includegraphics[width=0.8\textwidth]{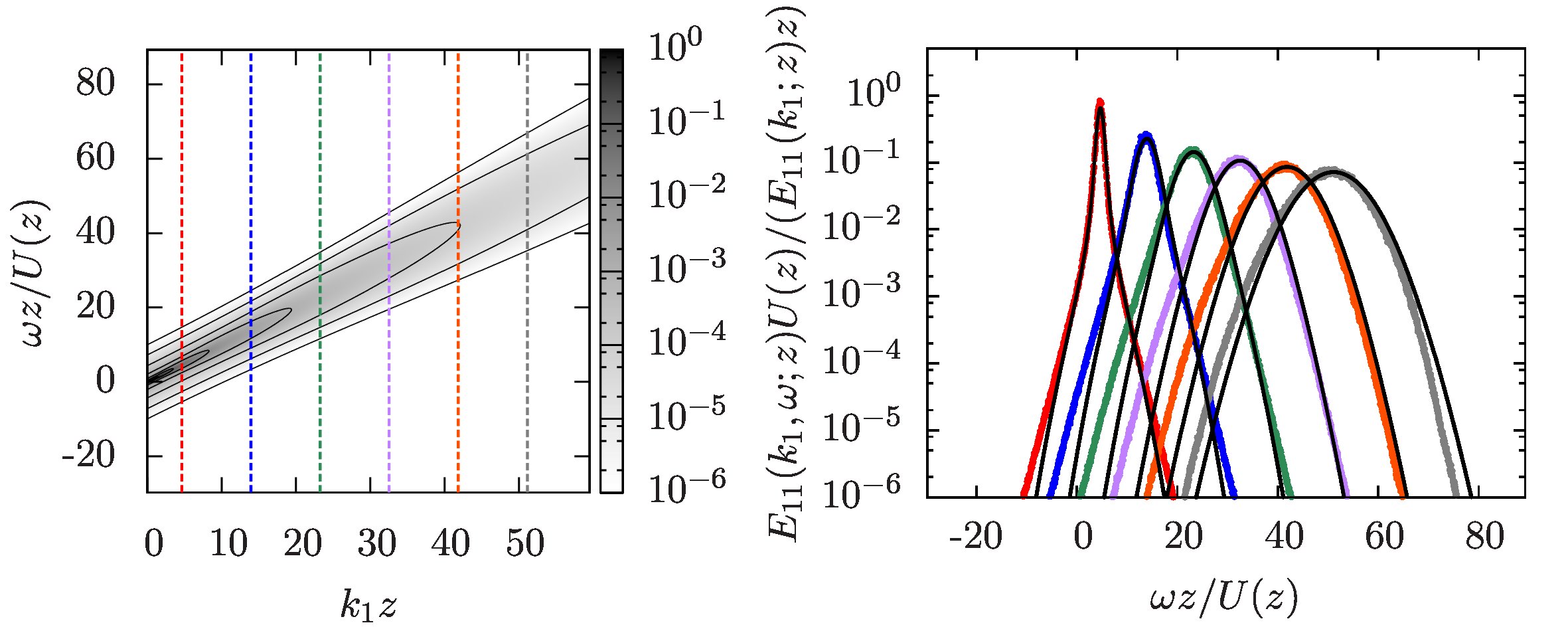} 
\end{center}
\caption{Streamwise wavenumber-frequency spectrum from the analytical model. From top to bottom $z/H \in \lbrace 0.0293,0.0684,0.104,0.154,0.232 \rbrace$. Left: model spectrum (see left panel of figure \ref{fig:stream_wf_spec_vs_model} for comparison with LES data), right: cuts showing frequency distributions for various wavenumbers (colored dots: LES data, black lines: model spectrum).}
\label{fig:stream_wf_spec_full_model}
\end{figure}

\begin{figure}
\begin{center}
    \includegraphics[width=0.8\textwidth]{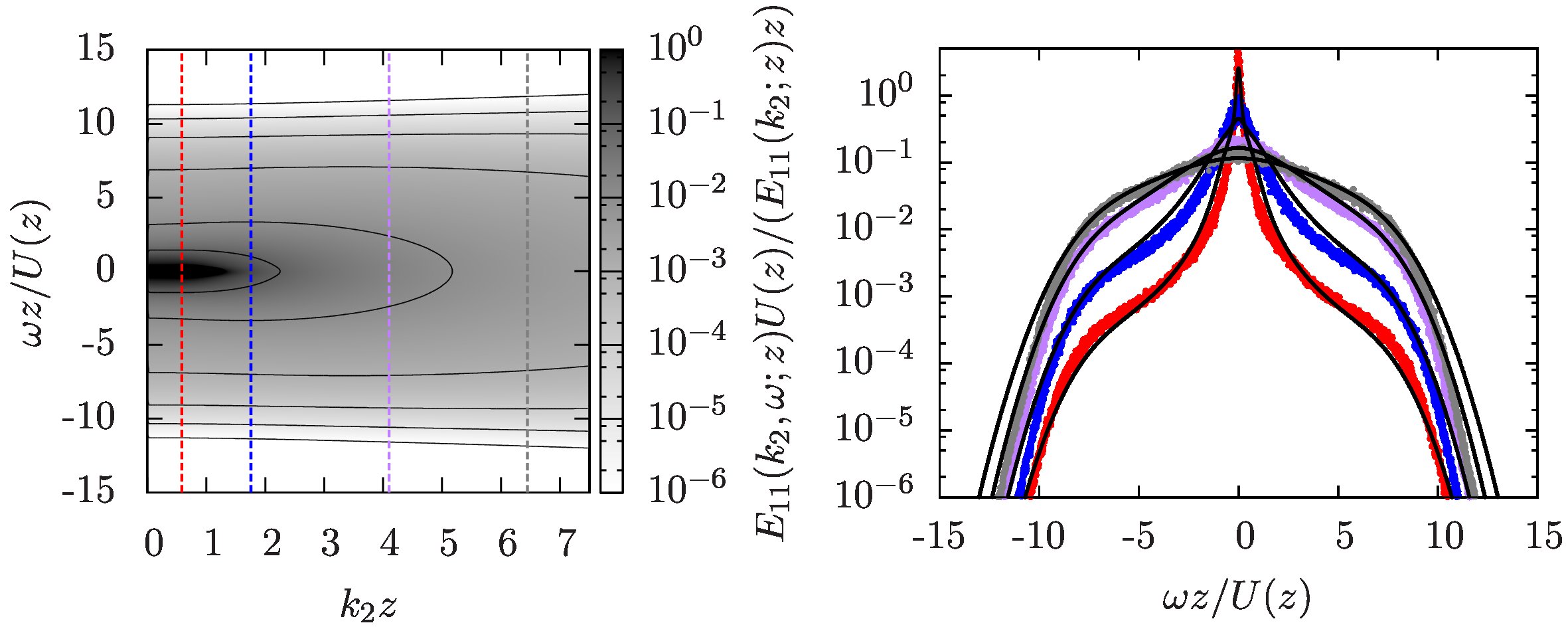}
    \includegraphics[width=0.8\textwidth]{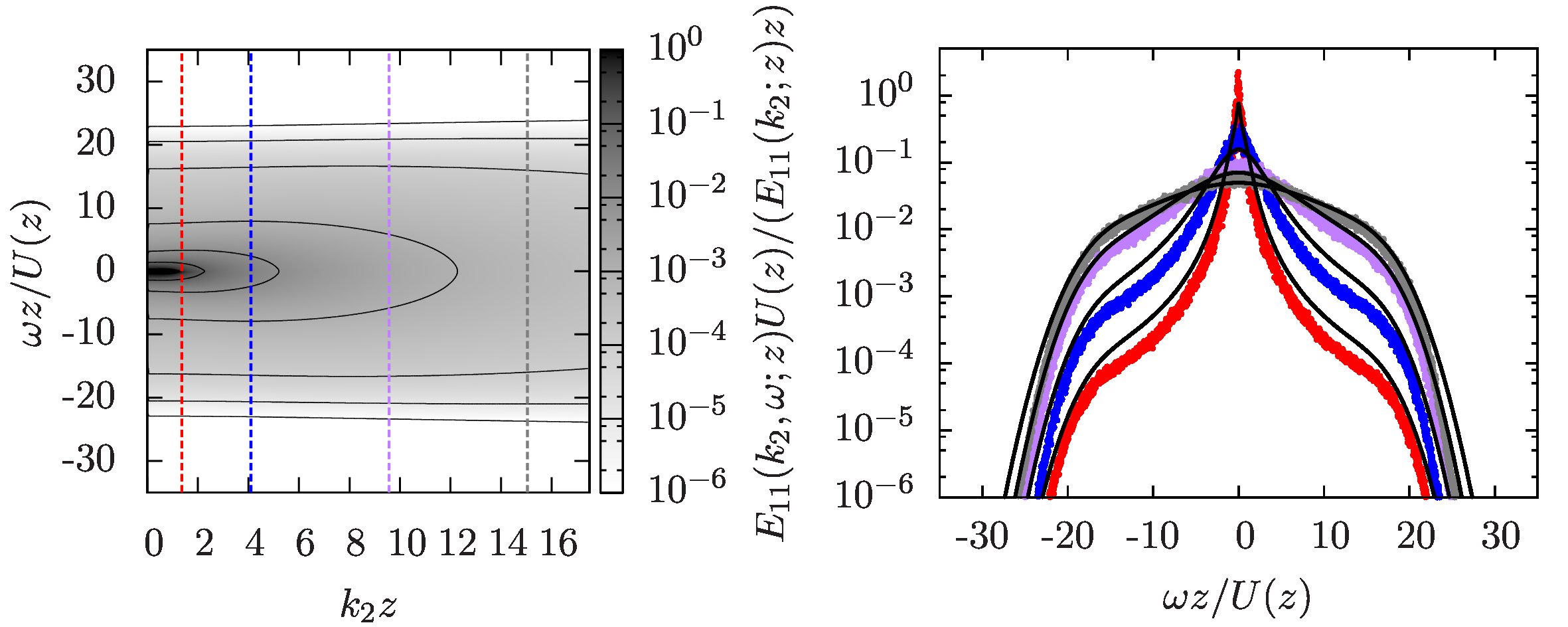}
    \includegraphics[width=0.8\textwidth]{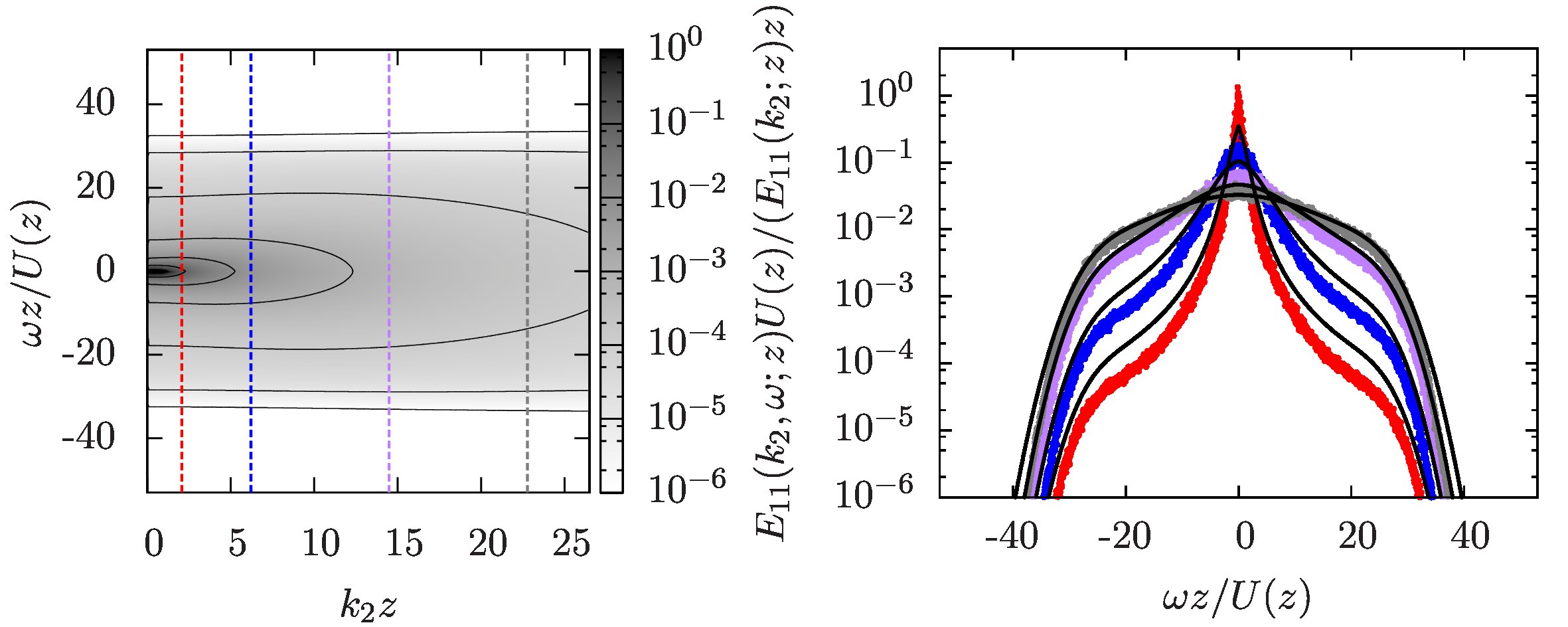}
    \includegraphics[width=0.8\textwidth]{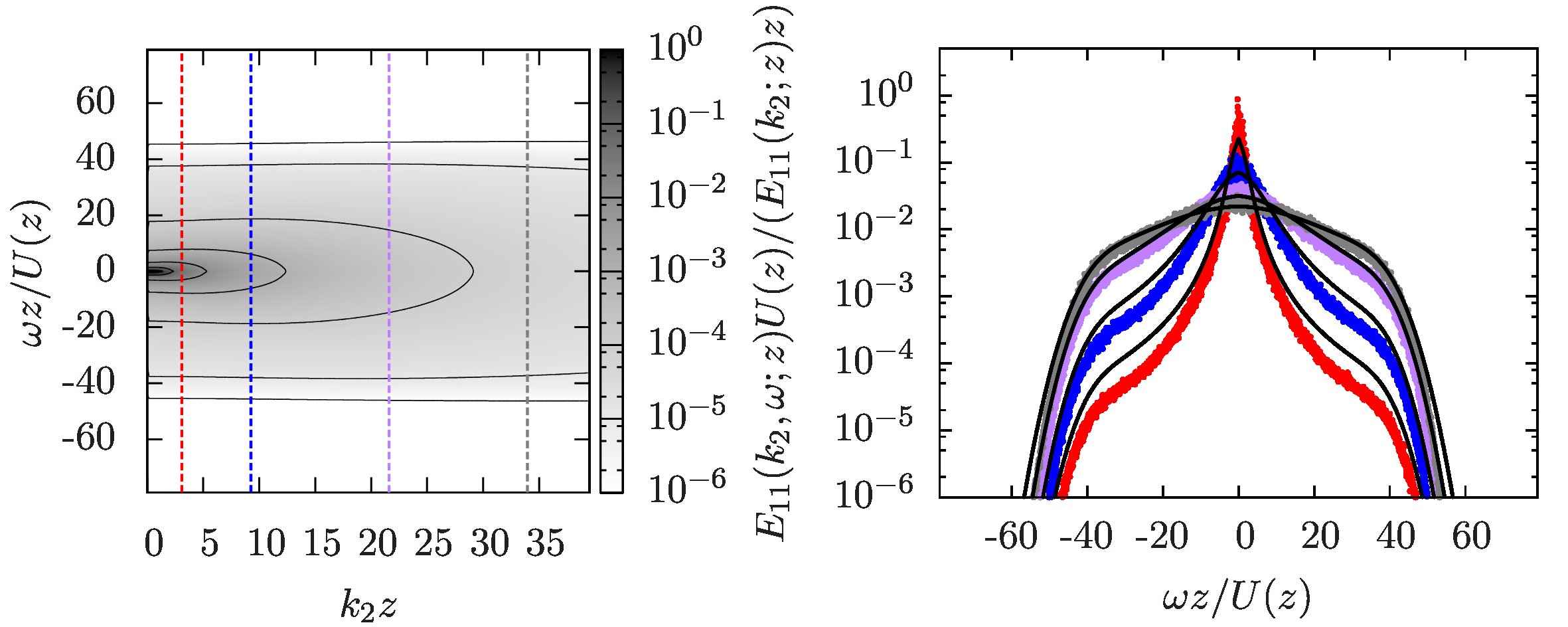}
    \includegraphics[width=0.8\textwidth]{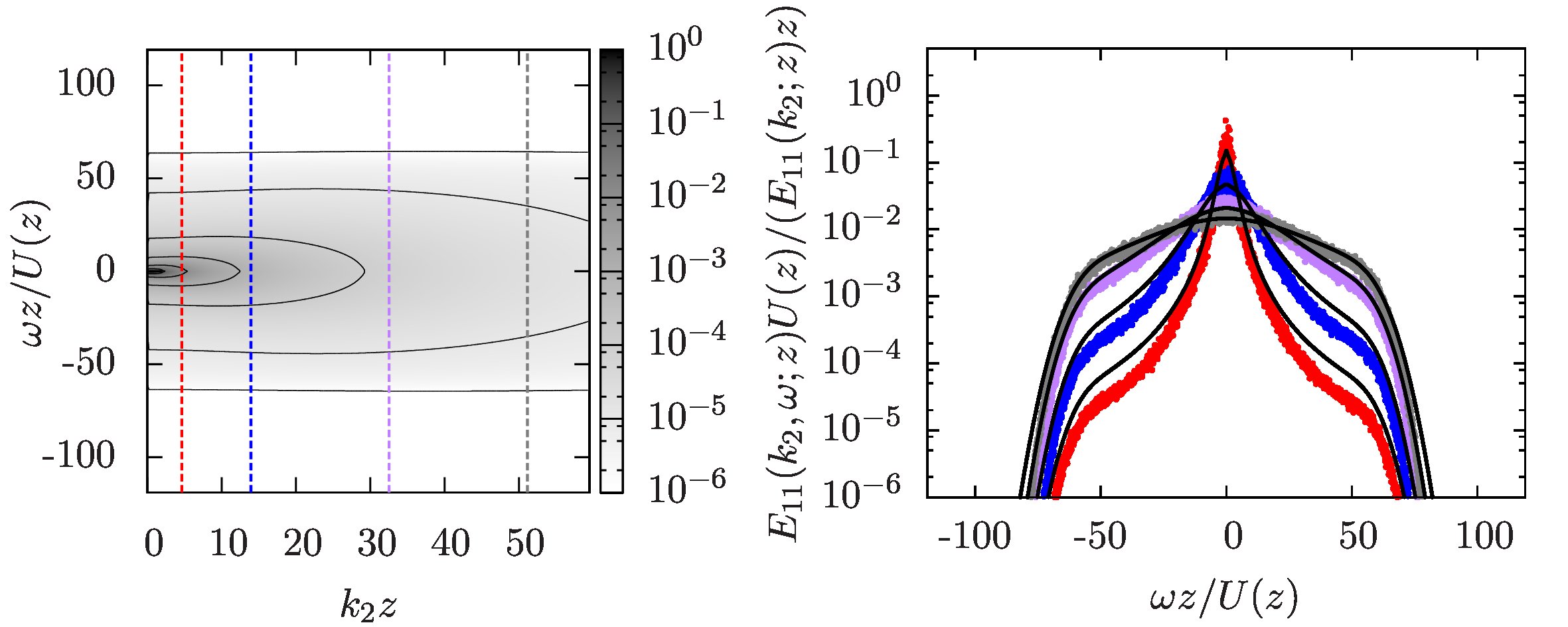}   
\end{center}
\caption{Spanwise wavenumber-frequency spectrum from the analytical model. From top to bottom $z/H \in \lbrace 0.0293,0.0684,0.104,0.154,0.232 \rbrace$. Left: model spectrum (see left panel of figure \ref{fig:span_wf_spec_vs_model} for comparison with LES data), right: cuts showing frequency distributions for various wavenumbers (colored dots: LES data, black lines: model spectrum).}
\label{fig:span_wf_spec_full_model}
\end{figure}

For application purposes, a model parameterization of \eqref{eq:wavenumberfrequencyspectrum} is of great use. Such a model was recently introduced in \cite{wilczek15jfm}, to which we refer the reader for further details. We here recapitulate the main ingredients of the model and present further benchmarks. In establishing a model for the joint wavenumber-frequency spectrum, we need to introduce a parameterization for the wavenumber spectrum $E_{11}(\bs k;z)$ and furthermore specify the mean velocity and random-sweeping effects for the frequency distribution.

To model the wavenumber spectrum, we smoothly blend between a low-wavenumber and a high-wavenumber contribution, $E^<_{11}$ and $E^>_{11}$, respectively:
\begin{equation}\label{eq:streamspanspecfullmodel}
  E_{11}(\bs k ; z) = \left[1 - \theta_{\alpha}\left( kz \right)\right] E^{<}_{11}(\bs k ; z) + \theta_{\alpha}\left( kz \right) E^{>}_{11}(\bs k ; z)  \, .
\end{equation}
$\theta_{\alpha}(x) = \frac{1}{2} \left( \tanh[ \alpha \log(x) ] +1 \right)$ is a smoothed-out step function with a parameter $\alpha$ controlling the steepness of the step (we here choose $\alpha=4$), and $k$ denotes the magnitude of the two-dimensional wave vector. For the high-wavenumber part of the spectrum we assume that the small-scale turbulence is homogeneous and isotropic with a Kolmogorov energy spectrum function $E\big(\tilde k \big) = C_{\mathrm{K}} \varepsilon^{2/3} \tilde k^{-5/3}$. Here, $C_{\mathrm{K}} \approx 1.6$ is the Kolmogorov contant, and $\varepsilon$ is the mean energy dissipation, estimated as $\varepsilon = u_*^3/(\kappa z)$. The magnitude of the three-dimensional wave vector is denoted by $\tilde k$. For the energy spectrum in the plane we obtain as a consequence
\begin{equation}\label{eq:smallscalespectrum}
  E^{>}_{11}(\bs k;z) = \int \! \mathrm{d}k_3 \, \frac{E\big(\tilde k \big)}{2 \pi {\tilde k}^2} \left( 1 - \frac{k_1^2 }{{\tilde k}^2} \right)   = \frac{\Gamma\left( \frac{1}{3} \right) C_{\mathrm{K}}}{5\sqrt{\pi}\Gamma\left( \frac{5}{6} \right)}  \varepsilon^{2/3} \left[ 1 - \frac{8}{11} \frac{k_1^2}{k^2} \right] k^{-8/3} \, .
\end{equation}
For the low wavenumbers, we blend between a constant spectrum and a $k_1^{-1}$ region with a Batchelor-type interpolation:
\begin{equation}\label{eq:largescalespectrum}
  E^{<}_{11}(\bs k;z) = D(z) z u_*^2 \left[ \left({1}/{H}\right)^{\beta}  + k_1^{\beta} \right]^{-1/\beta} \, .
\end{equation}
Here, $D(z)$ is a height-dependent amplitude determined numerically such that the integral over the spectrum matches the fluctuation variance given by the log law below. $\beta$ determines the sharpness of the transition, we here choose $\beta=4$.

The Doppler shift in the frequency distribution in \eqref{eq:wavenumberfrequencyspectrum} depends on the mean velocity, which in the log layer of wall-bounded flows takes the form
\begin{equation}\label{eq:loglawmean}
  U(z) = \frac{u_*}{\kappa} \log\left( \frac{z}{z_0} \right) \, .
\end{equation}
We assume $\kappa = 0.4$ for the von-K\'arm\'an constant. For the variance of the frequency distribution we need to model the term $\langle (\bs k \cdot \bs v)^2 \rangle$, for which we need to specify $\langle v_1^2 \rangle$ and $\langle v_2^2 \rangle$. According to the attached eddy hypothesis \citep{perry82jfm}, the streamwise velocity fluctuations also follow a log law, which leads us to the model parameterization
\begin{equation}\label{eq:loglawfluctuations}
  \left\langle v_1^2 \right\rangle = u_*^2 \left[ B - A\log\left( \frac{z}{H} \right) \right] .
\end{equation}
where $A$ denotes the ``Perry-Townsend" constant, and $B$ is a non-universal constant. In \cite{wilczek15jfm} we obtained the values from a model spectrum, which yields $A \approx 0.965$ and $B \approx 2.41$. For simplicity we assume $\langle v_2^2 \rangle = 0.41 \, \langle v_1^2 \rangle$, in which the constant of proportionality is motivated by our LES data. This concludes the model parameterization of the terms in \eqref{eq:wavenumberfrequencyspectrum}.

Figures \ref{fig:stream_wf_spec_full_model} and \ref{fig:span_wf_spec_full_model} show the streamwise and spanwise projections of the wavenumber-frequency spectrum evaluated from the analytical model. To this end, the model wavenumber spectrum has been discretized on a grid with the same resolution as the LES data, i.e. the analytical model spectrum exhibits the same high-wavenumber cutoff. For the streamwise projection a decent quantitative agreement is observed for all heights under consideration. The most pronounced differences can be seen in the tails of the frequency distribution at high wavenumbers. For the spanwise projections we observe some deviations in the width of the frequency distributions especially at low wavenumbers. As we have seen in the previous section that the random sweeping models captures the spectra quite accurately, these deviations most likely can be traced back to an imperfect modeling of the wavenumber spectrum.

\section{Effect of grid resolution}

\begin{figure}
\begin{center}
    \includegraphics[width=0.5\textwidth]{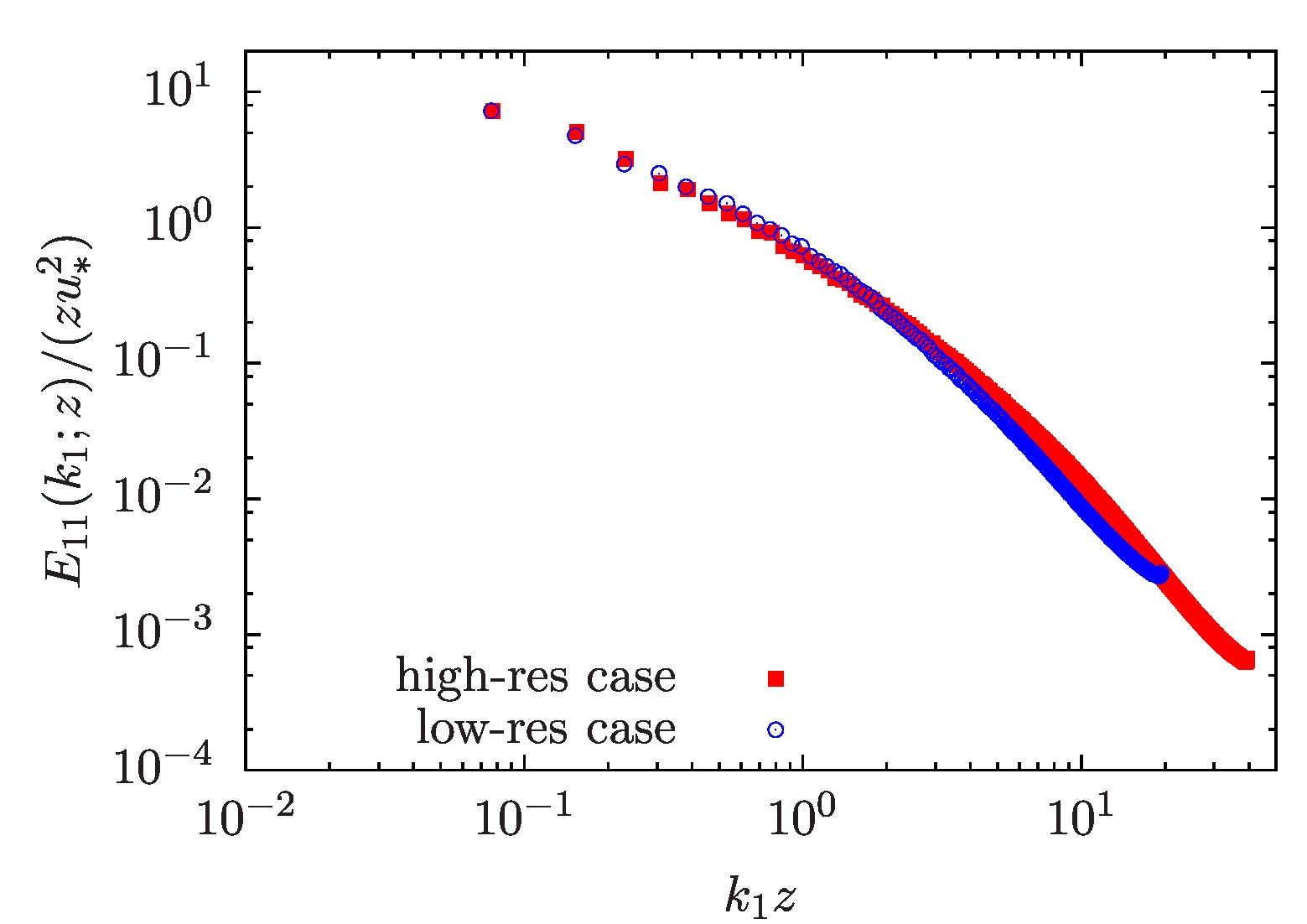}
\end{center}
\caption{Streamwise energy spectrum for the high-res case compared to the low-res case.}
\label{fig:spec_hilo_comp}
\end{figure}

For the evaluations presented so far we have used the data set with the highest spatial and temporal resolution suitable for this analysis that we have available (in the following referred to as the high-res case). Because the smallest spatial scales are modeled in LES, it is an interesting question to ask in what way our results are affected by the lack of small-scale resolution. A general assessment on accuracy of space-time correlations obtained from LES can be found in \cite{he04pof}. One of the conclusions of this study is that LES, compared to direct numerical simulations (DNS), tend to over-predict decorrelation time scales of turbulent flows.

As we cannot easily go beyond the resolution of the current data set, we take a lower-resolution simulation for comparison (in the following referred to as the low-res case although also this resolution can be considered reasonable for nowadays standards). The lower-resolution data set features the identical computational domain of $L_x/H \times L_y/H \times L_z/H = 4\pi \times 2\pi \times 1$, but it is resolved on a $512 \times 256 \times 128$ grid, corresponding to case C2 in \cite{ste14d}. We collected data for $4.1 \times 10^4$ time steps at a fixed time step of $5.5 \times10^{-5} H/u_*$.

Both data sets display a similar mean velocity ($U \approx 19.2 u_*$ for the low-res case compared to $U \approx 19.0 u_*$ for the high-res case) as well as comparable fluctuations ($\langle v_1^2 \rangle \approx 4.37 u_*^2$ and $\langle v_2^2 \rangle \approx 1.60 u_*^2$ for the low-res case compared to $\langle v_1^2 \rangle \approx 4.33 u_*^2$ and $\langle v_2^2 \rangle \approx 1.70 u_*^2$ for the high-res case). In the following we are considering spectra at a height of $z/H \approx 0.152$ in the low-res case and $z/H \approx 0.154$ in the high-res case.

The effect of the different spatial resolution of the two data sets is demonstrated with a plot of the streamwise wavenumber spectra of the streamwise velocity fluctuations as presented in figure \ref{fig:spec_hilo_comp}. The most striking difference is the earlier cutoff in the low-res case, although there are some differences also in the shape of the spectra.

\begin{figure}
\begin{center}
    \includegraphics[width=0.58\textwidth]{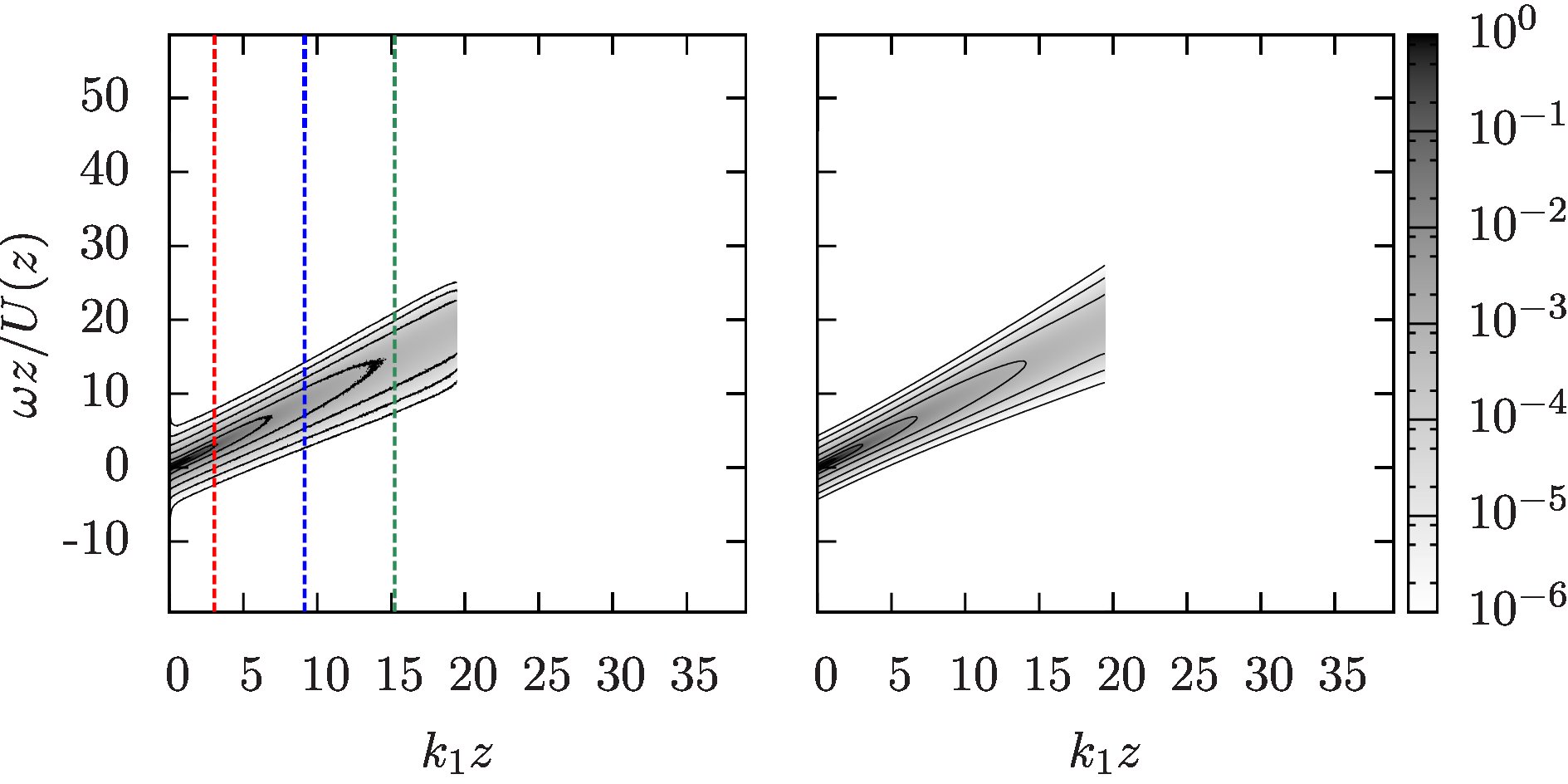}
    \includegraphics[width=0.41\textwidth]{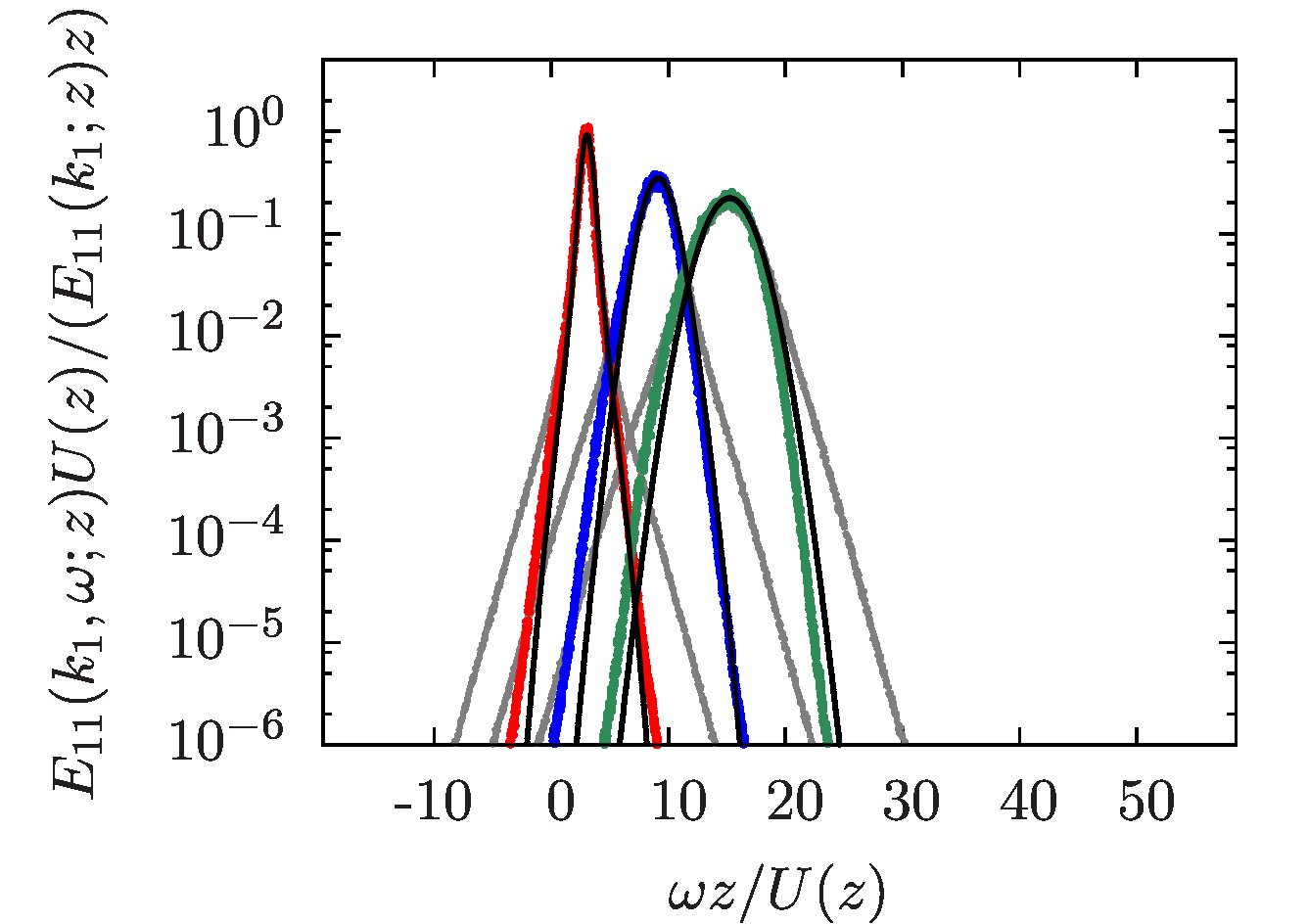}
    \includegraphics[width=0.58\textwidth]{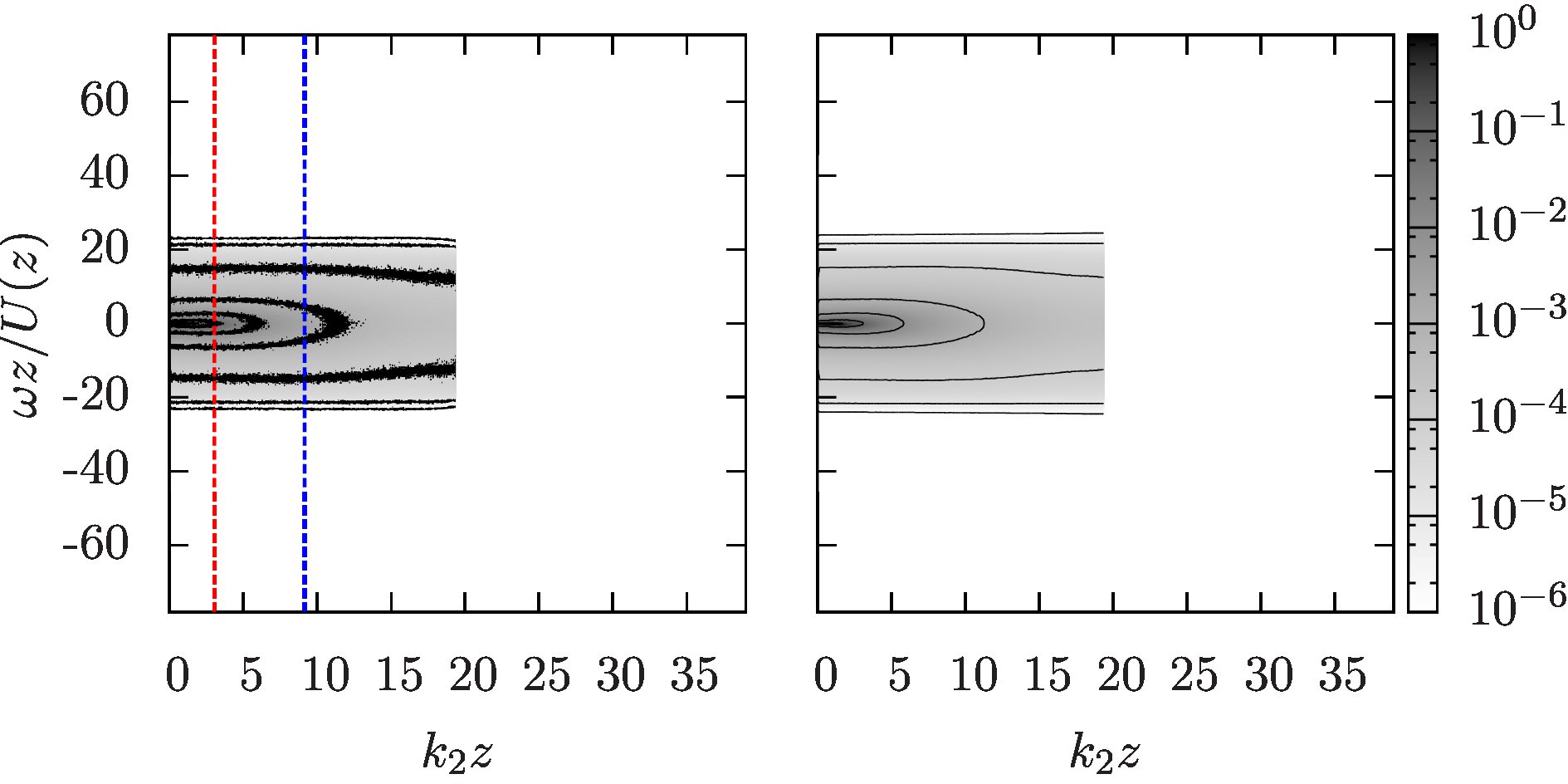}
    \includegraphics[width=0.41\textwidth]{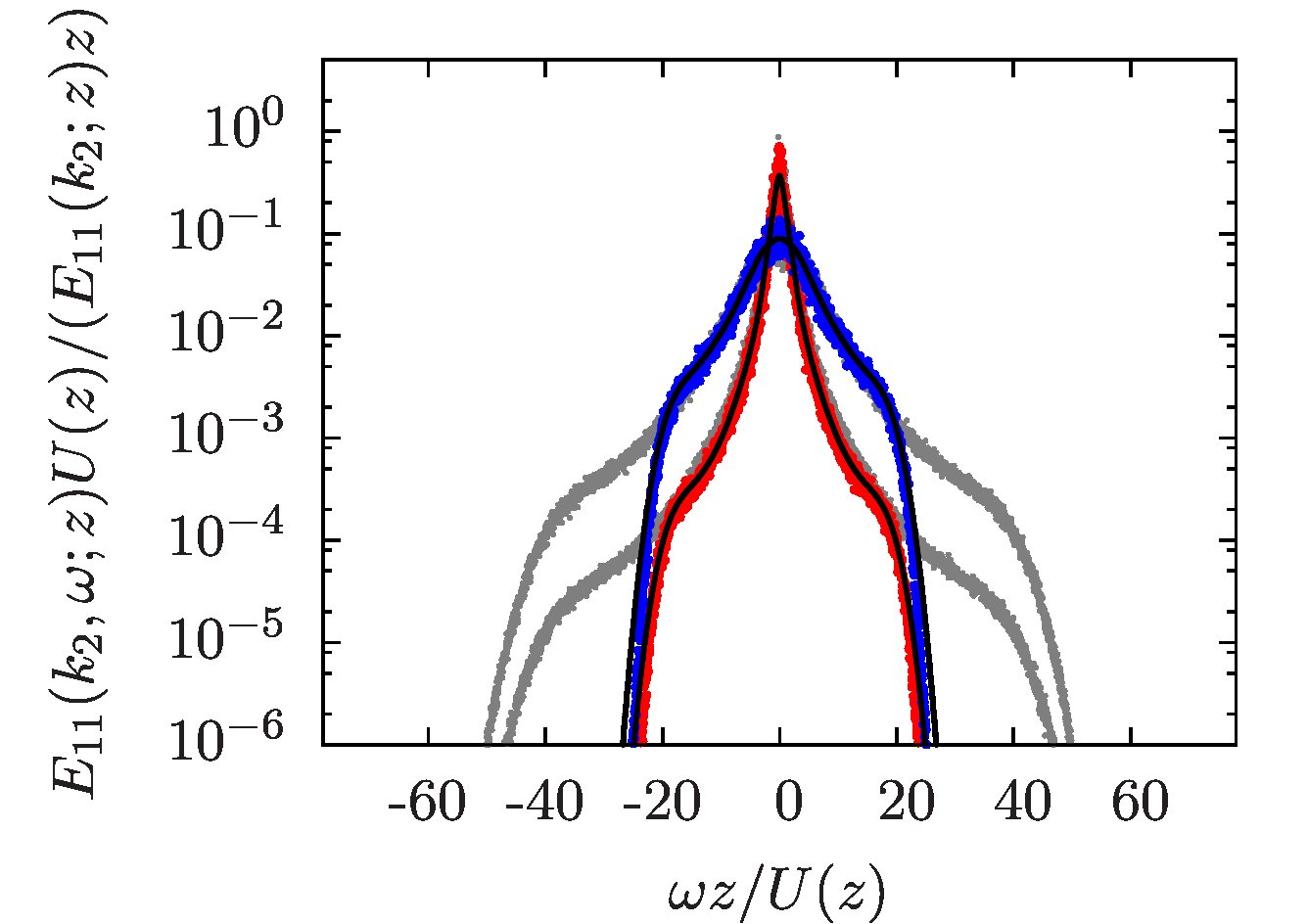}
\end{center}
\caption{Streamwise wavenumber-frequency spectrum (upper panel) and spanwise wavenumber-frequency spectrum (lower panel) for the streamwise velocity component in the low-res case at height $z/H \approx 0.152$. Left: LES data, middle: prediction based on random sweeping hypothesis, right: cuts comparing LES data (colored dots) and random sweeping model (black lines). The gray curves correspond to the frequency distributions in the high-res case. To allow for a direct comparison with the spectra from the high-res case at $z/H \approx 0.154$ in figures \ref{fig:stream_wf_spec_vs_model} and \ref{fig:span_wf_spec_vs_model}, respectively, the axis ranges have been chosen identically.}
\label{fig:wf_spec_vs_model_lores}
\end{figure}

The wavenumber-frequency spectra $E_{11}(k_1,\omega;z)$ as well as $E_{11}(k_2,\omega;z)$ of the lower-res case are presented in figure \ref{fig:wf_spec_vs_model_lores}. Besides the limited wavenumber range the spectra also appear narrower with respect to the frequency range. The right panel shows that, while the cores of the frequency distributions collapse in good approximation, the tails of the frequency distributions are reduced when compared to the results from the high resolution case in figures \ref{fig:stream_wf_spec_vs_model} and \ref{fig:span_wf_spec_vs_model} (and the frequency distributions of the high-res case also shown in figure \ref{fig:wf_spec_vs_model_lores}). As a consequence, the Doppler broadening is lower in the low-res case. The differences can be understood in the framework of the random sweeping hypothesis. According to \eqref{eq:wavenumberfrequencyspectrum} the variance of the frequency distribution depends quadratically on the wavenumbers. One consequence is that lower-resolution simulations lack some of the high-wavenumber contributions at which the wavenumber-frequency spectrum exhibits a considerable Doppler broadening. These contribution are also missing upon projection, for example, to the $k_1$-$\omega$-plane, see \eqref{eq:projection}. Because the low-$k_1$ range contains also high-$k_2$ contributions, also this range exhibits a reduced Doppler broadening. The same argument, of course, also applies to the projection to the $k_2$-$\omega$-plane. Because a reduced Doppler broadening is related to a decreased decorrelation, these findings appear consistent with the ones presented in \cite{he04pof}.

\begin{figure}
\begin{center}
    \includegraphics[width=0.8\textwidth]{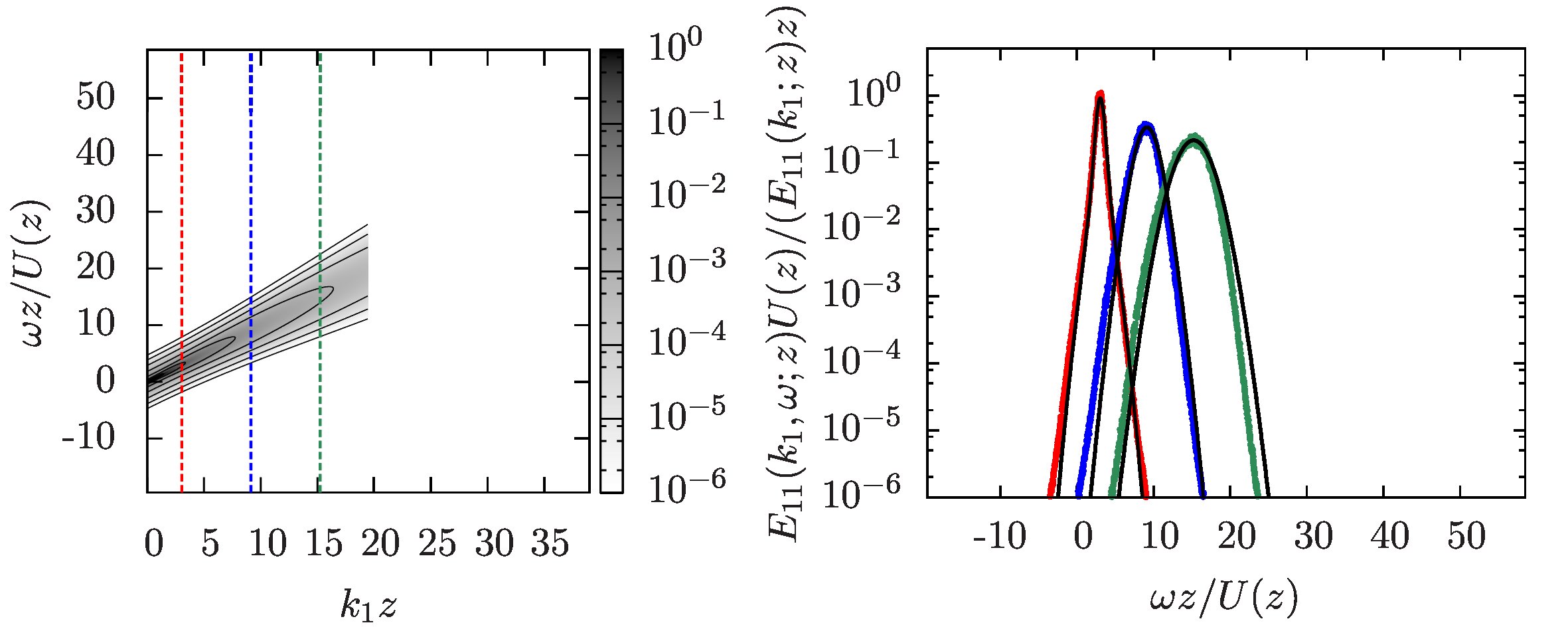}
\end{center}
\caption{Streamwise wavenumber-frequency spectrum from the analytical model for the low-res case at $z/H \approx 0.152$. Left: model spectrum (see upper left panel of figure \ref{fig:wf_spec_vs_model_lores} for comparison with LES data), right: cuts showing frequency distributions for various wavenumbers (colored dots: LES data, black lines: model spectrum).}
\label{fig:stream_wf_spec_full_model_lores}
\end{figure}

Figure \ref{fig:stream_wf_spec_full_model_lores} exemplarily shows a comparison to the analytical model spectrum, projected to the $k_1$-$\omega$-plane. As in the high-res case the spectrum has been discretized on a grid matching the LES resolution. The remaining parameters have not been changed. As in the high-res case we observe satisfactory agreement.

These comparisons highlight that the quantitative details of the wavenumber-frequency spectra, specifically the magnitude of the Doppler broadening, depend on the resolution of the LES whereas the qualitative features remain.

\section{Summary}

We have presented results obtained from large-eddy simulations on the height-dependence of spatio-temporal spectra of the streamwise velocity component in the logarithmic layer of wall-bounded flows. These observations reveal a considerable Doppler shift and Doppler broadening of frequencies, which can be explained by advection of small-scale fluctuations by a mean flow and large-scale random sweeping effects. The strength of Doppler shift and broadening thus varies with height.

In an earlier publication \cite{wilczek15jfm} these effects have been taken as the main ingredients for a simple advection model, which we have tested  further here. We specifically focused on the height-dependence of the spectra in the logarithmic layer and presented projections to the $k_1$-$\omega$ plane as well as to the $k_2$-$\omega$ plane. To extend the model beyond the logarithmic range, models like the one proposed by Marusic et al.~\cite{marusic10sci} predicting velocity and wall stress fluctuations very close to the wall including non-trivial correlations across wall normal positions could turn out to be useful.

We generally find that the spectra obtained under the random sweeping hypothesis with mean flow are in very good agreement, which motivated us to take the predictions of this model as a starting point for an analytical parameterization of the full spatio-temporal spectra.

We then presented additional benchmarks on this analytical model. While we find good qualitative agreement, some quantitative differences are observed, rooted mainly in an imperfect modeling of the wavenumber part of the spectrum. Furthermore, we evaluated the influence of LES grid resolution showing that insufficient resolution tends to under-predict Doppler broadening.

As a future application, we plan to extend this model to predict spatio-temporal properties of fluctuations in wind farms, a topic of ongoing research.
\\
\\
{\it Acknowledgments:} 
M.W. was supported by DFG funding WI 3544/2-1 and WI 3544/3-1, R.J.A.M.S. by the `Fellowships for Young Energy Scientists' (YES!) of FOM, and C.M. by US National Science Foundation grant \#IIA-1243482 (the WINDINSPIRE project). Computations were performed with SURFsara resources, i.e. the Cartesius and Lisa clusters. This work was also supported by the use of the Extreme Science and Engineering Discovery Environment (XSEDE), which is supported by National Science Foundation grant number OCI-1053575.

\end{document}